\documentclass[aps,physrev,preprint,superscriptaddress]{revtex4-2}

\usepackage{graphicx}
\usepackage{dcolumn}
\usepackage{bm}
\usepackage{amsmath}%
\usepackage[dvipsnames]{xcolor}

\begin{document}


\title{Near-Infrared noise in intense electron bunches}

\author{Sergei Kladov}
\email{Contact author: kladov@uchicago.edu}
\affiliation{University of Chicago, Chicago, USA}
\author{Sergei Nagaitsev}
\affiliation{Brookhaven National Laboratory, Upton, USA}
\author{Young-Kee Kim}
\affiliation{University of Chicago, Chicago, USA}
\author{Daniel R. Broemmelsiek}
\affiliation{Fermi National Accelerator Laboratory, Batavia, USA}
\author{Zhirong Huang}
\affiliation{SLAC National Accelerator Laboratory, Menlo Park, USA}
\author{Jonathan Jarvis}
\author{Alex H. Lumpkin}
\author{Jinhao Ruan}
\author{Andrea Saewert}
\author{Randy M. Thurman-Keup}
\affiliation{Fermi National Accelerator Laboratory, Batavia, USA}


\date{\today}

\begin{abstract}
This article investigates electron bunch density fluctuations in the 1 -- 10 $\mu m$ wavelength range, focusing on their impact on coherent electron cooling (CEC) in hadron storage rings. In this study, we compare the shot-noise model with experimental observations using bandwidth-filtered near-infrared (NIR) optical transition radiation (OTR) photodiode signals with the transverse bunch size being much larger than the OTR wavelength of interest. The relativistic electron bunch ($\gamma \approx 50$) parameters are close to those proposed for the Coherent Electron Cooler in the Electron-Ion Collider (EIC) project.  Preliminary feasibility experiments were conducted, and the noise levels with Fano factors are presented, supported by particle tracking. Longitudinal space-charge-induced microbunching for the chicane-compressed bunch was also observed with coherent OTR enhancements up to 100 times, which served as an additional calibration of the measurement method.
\end{abstract}


\maketitle

\section{Introduction}
The significance of bunch quality in relativistic electron bunches became increasingly prominent as the allowed density fluctuations, often called "noise", approached the Poisson statistics level (shot noise \cite{Schottky_1918, Schottky_original}). Various applications where it is important include high-energy hadron cooling systems \cite{stupakov2019moreNoiseInvestigations,bergan2024electronNoiseInCEC} and free-electron lasers (FELs) \cite{huang2007reviewSASEandSelfSeeded,saldin2002seededFELnoise}.

The luminosity in colliders depends mainly on bunch intensities and emittances, which in turn depend on many factors and processes, such as intrabeam scattering \cite{IBS_ref}. To maintain hadron bunch quality and high brightness, bunch cooling can be implemented in storage rings. Currently, two methods are employed: electron cooling \cite{derbenev2017ElectronCoolingTheory} and stochastic cooling \cite{vanderMeer:1972sf, van1995stochastic}. While electron cooling has been extensively studied and widely implemented, its cooling rate drops significantly with energy increase \cite{Lebedev_2024}, with the highest hadron energy cooled by this method being 8.9 GeV \cite{ECool_PhysRevLett.96.044801}.

\begin{figure}[htb]
\centering
\includegraphics[width=1\columnwidth]{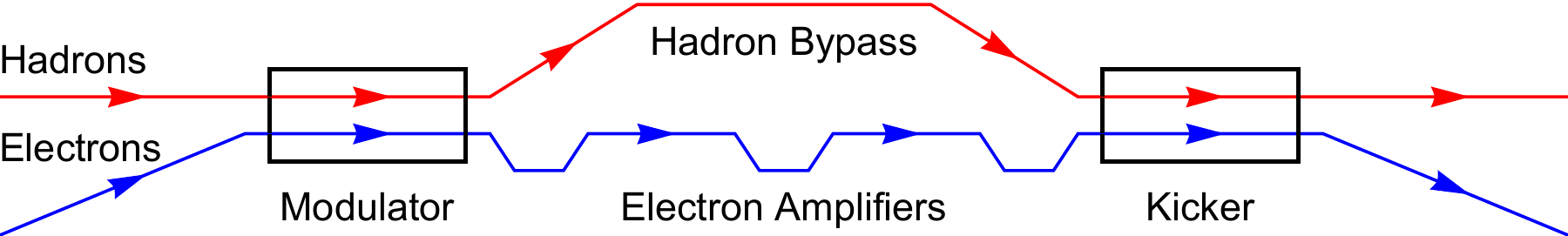}
\caption{\label{CECscheme} Coherent Electron Cooling (CEC) scheme.}
\end{figure}

Stochastic cooling functions as a broad-band feedback system by damping hadron bunch density fluctuations, detected by a 'pick-up' device, amplified, and then applied to a 'kicker' device at an optimal signal phase. Its cooling rate is not strongly dependent on bunch energy, and is also proportional to the device's operational frequency \cite{Lebedev_2024}, making Optical Stochastic Cooling (OSC) \cite{Lebedev_2021}, for example, orders of magnitude more effective than microwave stochastic cooling schemes and even more so than electron cooling at high energies \cite{OSCZolotarev}.  Like the OSC method, Coherent Electron Cooling (CEC) is a variation of transit-time stochastic cooling, where the same electron bunch acts as both the 'pick-up' modulator and the kicker \cite{Derbenev_CeC_1992, LitvinenkoCEC, CECwithMicrobunchingAsAmplifier}. Initially, the electron bunch receives a longitudinal density imprint, shaped by the hadron bunch \cite{nagaitsev:ipac2021-wepab273}; this kick is then amplified and transformed into longitudinal density modulation by a chicane, and finally, the hadron bunch is kicked by the modulated electron bunch with a specific phase shift, achieving cooling. The basic system scheme, to be built at the Electron Ion Collider (EIC), is depicted in Fig.~\ref{CECscheme} \cite{eicTechReport}.

Additional electron bunch density modulations introduce extra diffusion in the hadron bunch, counteracting cooling \cite{stupakov2018coolingShotNoiseEffect,stupakov2019moreNoiseInvestigations,bergan2024electronNoiseInCEC}. The evolution of the longitudinal hadron bunch emittance, $\varepsilon_L$, is determined by \cite{nagaitsev:ipac2021-wepab273}
\begin{equation}
    \frac{1}{\varepsilon_L} \frac{d\varepsilon_L}{dt} = -\frac{1}{T_{\text{cool}}} + 
    \frac{D}{\varepsilon_L},
\end{equation}
where $\varepsilon_L/D$ is the characteristic diffusion time, $T_{\text{diff}}$, and $1/T_{\text{cool}}$ is the optimized cooling rate with no noise taken into account. The diffusion coefficient $D$ contains contributions from both the hadron bunch and the electron bunch density fluctuations. The hadron bunch density fluctuations are Poisson-distributed (shot noise), and their contribution is well studied in \cite{stupakov2019moreNoiseInvestigations,bergan2024electronNoiseInCEC}. Following the notations from \cite{stupakov2019moreNoiseInvestigations}, it can be described by the ratio of the rates:

\begin{equation}
    \frac{T_{\text{cool}}}{T_{\text{diff}_h}}=r_1,
\end{equation}

where $T_{\text{diff}_h}$ is the characteristic time of the diffusion caused by the shot noise in the hadron bunch, and $r_1 = 0.04$ for EIC parameters \cite{eicTechReport}.

Therefore, we focus our attention on the electron bunch noise only. The shot level noise taken into account at the entrance to the amplifier section is treated similarly and results in a similar expression:

\begin{equation}
    \frac{T_{\text{cool}}}{T_{\text{diff}_e}}=r_2,
\end{equation}

where $r_2 = 0.02$ for EIC parameters. Besides the base effect $r_2$, the electron noise contribution contains an effect in the modulator, as well as the interference term \cite{bergan2024electronNoiseInCEC}. Following Appendix \ref{ap:noise_and_EIC_CEC}, we neglect these terms, and also take into account the density fluctuations that are different from the shot noise:

\begin{equation}
    \frac{T_{\text{cool}}}{T_{\text{diff}_e}}=r_2 \frac{\int_{-\infty}^{\infty}{\left|Z_{e,2}(\omega)\right|^2\left|\delta \rho_e(\omega)\right|^2 d\omega}}{\int_{-\infty}^{\infty}{\left|Z_{e,2}(\omega)\right|^2 \frac{1}{N} d\omega}},
    \label{DdifftoDcool1}
\end{equation}


where $r_2 = 0.02$ is the ratio for a shot-noise (quiet) bunch, $N$ is the number of electrons in the bunch, and $Z_{e,2}(\omega)$ is the total impedance of the CEC amplifier and kicker sections for pre-existing electron density modulations $\delta \rho_e(\omega)$.

The modulations are defined as the difference between the expected smooth unperturbed distribution function $\rho_0(\omega)$ and the actual distribution function, composed of $N$ electrons, arriving at times $t_j$, represented by Dirac delta-functions (Fourier transformed):

\begin{equation}
    \delta \rho_e(\omega) = \rho_0(\omega) - \rho(\omega) = \rho_0(\omega) - \frac{1}{N}\sum_{j=1}^{N}{\text{e}^{-i \omega t_j}},
\end{equation}

where $j$ is the particle index and the distribution functions are normalized on 1.

\begin{figure}[htb]
\centering
\includegraphics[width=0.48\columnwidth]{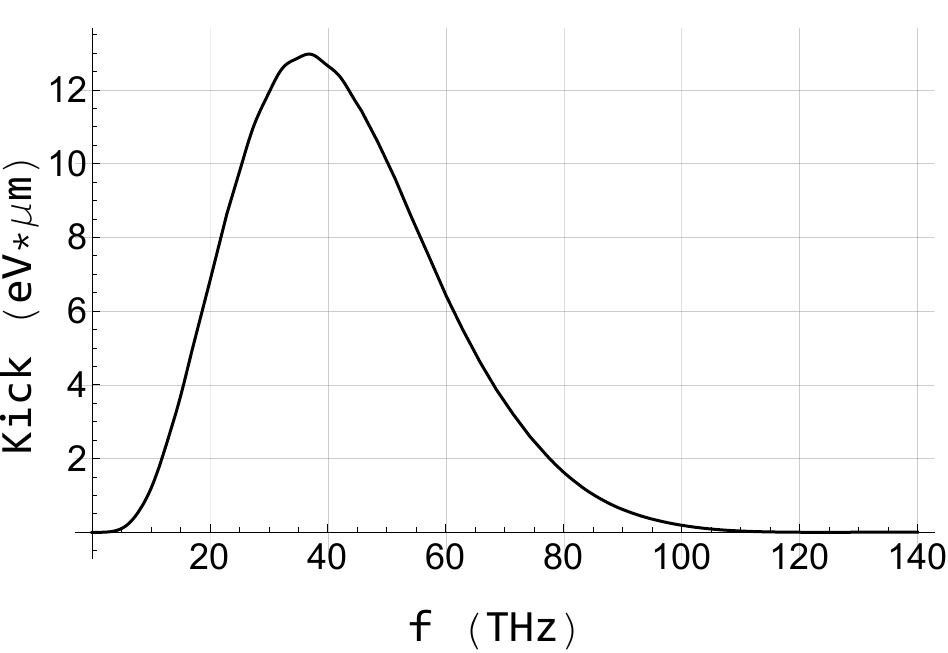}
\quad
\includegraphics[width=0.48\columnwidth]{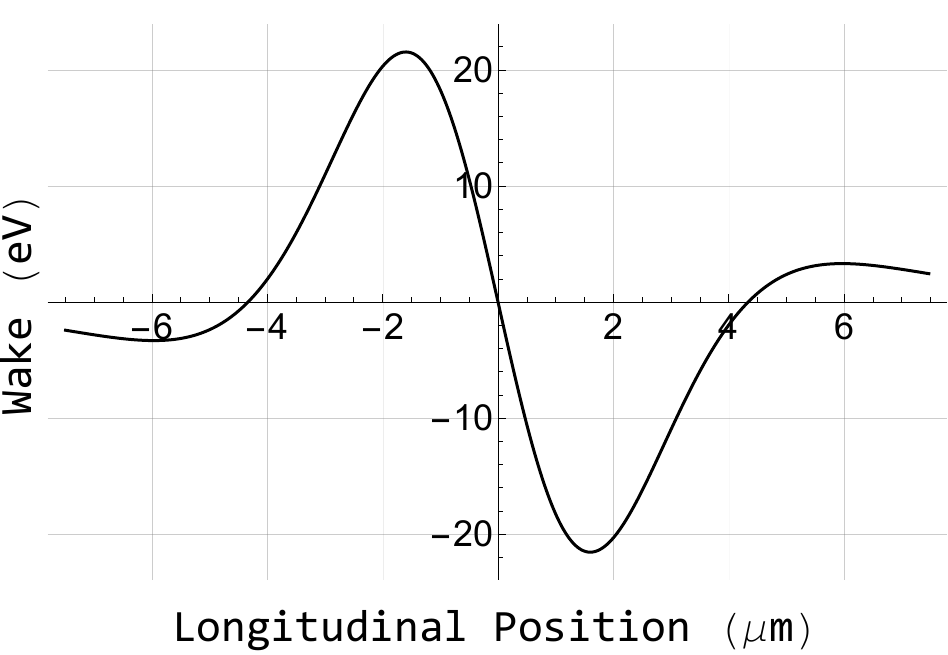}
\caption{\label{CECimpedance} Total impedance of the Coherent Electron Cooling (CEC) amplifier and kicker sections at EIC in frequency ($f = \frac{1}{2\pi}\omega$) domain and the wake function in spacial ($z = c t$) domain. The characteristic length is 1~$\mu$m.}
\end{figure}

The impedance (Fig~\ref{CECimpedance}, left), which is the Fourier transform of the temporal ‘kick’, applied to hadrons (Fig~\ref{CECimpedance}, right), is a convenient tool, broadly used by CEC community \cite{bergan2024electronNoiseInCEC, stupakov2019moreNoiseInvestigations}. If wakes are used, Eq. \ref{DdifftoDcool1} transforms to a heavier expression without an obvious connection to experimental measurements. Note that the system's response characteristic length is about 1~$\mu$m.

The same noise is also relevant in FELs. In self-amplified spontaneous emission (SASE) and self-seeded FELs the microwave instability starts from the initial noise in the bunch, leading to the bunch microbunching and resulting in coherent radiation. Hence, the initial noise within the FEL bandwidth plays a beneficial role \cite{huang2007reviewSASEandSelfSeeded}. Conversely, in externally seeded FELs, such noise disrupts the seed signal, and reducing noise at the initial seed wavelength can lower the seed laser power requirements \cite{saldin2002seededFELnoise,stupakov2010seededFELnoise}.

There have been several proposals in the past to suppress the noise in the bunch in the frequency range of interest in order to optimize the cooling effects~\cite{LitvinenkoFEL09}. However, before suppressing the noise it is necessary to understand its amplitude and spectral decomposition, as well as its origin. Although the noise exceeding shot level has been already observed in linear electron accelerators without external wavelength modulation \cite{gover2012noiseLevelDecreaseInDriftSection}, there is no comprehensive analysis and simulation support for low-level fluctuations, especially in the range of bunch parameters similar to EIC CEC.

In this paper we present measurements of the electron bunch density noise level in the 0.5 – 3 $\mu$m wavelength range, and compare them with analytical and tracking predictions. We show that if the bunch has parameters similar to ones in the proposed EIC CEC scheme, it does not have  major fluctuations in the NIR region, while the noise becomes much stronger if the bunch is longitudinally compressed.

The article is organized as follows. In sections \ref{sec:methods}, \ref{sec:experimentalSetup} we give an overview of the experiment: general idea and technical details. It is followed by section \ref{sec:theory}, devoted to provide a theoretical basis for predicting and analyzing the experimental data. This data is presented in section \ref{sec:exp_results}, alongside with a discussion. Next, the simulations and the tracking results are briefly described in section \ref{sec:simulations}. Finally, section \ref{sec:comparison} contains a discussion about the compressed beam and the elevated noise, together with a list of addressed issues and effects.

The electron bunch noise influence on the cooling rate at EIC CEC is described in appendix \ref{ap:noise_and_EIC_CEC}. Appendix \ref{ap:optical_channel} provides a detailed study of the optical channel, necessary for experimental data understanding, and appendix \ref{ap:MainNEBparameters} summarizes the main numerical values used in the article.

\section{Methods}
\label{sec:methods}

Density fluctuations in electron bunches are traditionally measured by means of Transition Radiation (TR). The connection between the density fluctuations in the bunch and the TR signal has been well established in FELs, where the space charge impedance in a long linear accelerator, as well as strong compression to a high peak current, often result in bunch micro-structures \cite{HUANG2018182} that strongly amplify the TR in the optical range of frequencies, known as coherent optical transition radiation (COTR) \cite{LoosLCLSCOTR}.

The OTR energy per unit frequency $\omega$ per unit solid angle, radiated by a bunch of charged particles, traversing an ideal conducting plane, can be approximated as (see sec. \ref{sec:theory}):

\begin{equation}
  \frac{d^2W}{d\omega d\Omega} = \frac{d^2W_1}{d\omega d\Omega} N^2\left|\rho(\omega)\right|^2,
\end{equation}

where $\rho(\omega)$ is the Fourier spectrum of the normalized longitudinal bunch density distribution function ($\rho(0)=1$), and the spectrum for one particle is

\begin{equation}
    \frac{d^2W_1}{d\omega d\Omega}=\frac{Z_0 q^2}{4 \pi^3} \frac{\beta^2 {\sin^2 \theta}}{{(1-\beta^2 {\cos^2 \theta})}^2} \approx 
    \frac{Z_0 q^2}{4 \pi^3} \frac{\theta^2}{{(\gamma^{-2}+  \theta^2)}^2},
    \label{otr_energy}
\end{equation}

where $\beta = v/c$ is the particle velocity normalized by speed of light, $Z_0 \approx 377 ~ \Omega$ is the impedance of free space, $q$ is the single particle charge, and the relativistic factor is $\gamma \gg 1$. In this approximation we assumed that the bunch size is small \(\sigma_\perp\ll\lambda\). In this ultra-relativistic case most of the OTR radiation is concentrated in a cone with $|\theta| < 1/\gamma$. Therefore, the photons can be efficiently transported from the foil to the detection system. It is also seen from the expression that an elevated level of noise results in a higher energy of the radiation, possibly with a spiky spectrum, and faster signal scaling with charge (e.g. quadratic). Assuming a bunch charge of 1~nC, the radiation energy $W$ per quiet bunch (shot noise, $\left|\rho(\omega)\right|^2 = 1/N$) in the band $\Delta \lambda=100~\text{nm}$, $\lambda_0=1\,\mu$m at 25~MeV is $W=2.4$~pJ.

\section{Experimental setup}
\label{sec:experimentalSetup}

The experiments were performed at Fermilab Accelerator Science and Technology (FAST) facility as it can provide electron bunches with similar bunch parameters as in the EIC CEC concept. The comparison is presented in Table~\ref{tab:fast_eic_compare}. The most important parameters for this project are summarized in Appendix \ref{ap:MainNEBparameters}.

\begin{table}[!hbt]
   \caption{\label{tab:fast_eic_compare}%
   FAST and Proposed CEC bunch Parameters.}
   \begin{ruledtabular}
   \begin{tabular}{lcr}
       \textbf{Parameter} & \textbf{FAST}                      & \textbf{EIC} \\
       \colrule
           $E_p$, GeV &  & 100 - 275        \\ 
           $E_e$, MeV & 40 - 300 & 50 - 150        \\ 
           Bunch charge, nC & 0 - 3 & 1        \\ 
           $\varepsilon$ (rms, norm), $\mu$m  & 3 (at 1 nC) & 2.8        \\ 
           Bunch length, mm & 0.3 - 10 & 12 - 8        \\ 
           Drift section, m & 80 & 100 \\
   \end{tabular}
   \end{ruledtabular}
\end{table}

\begin{figure}[htb]
\centering 
\includegraphics[width=1\textwidth]{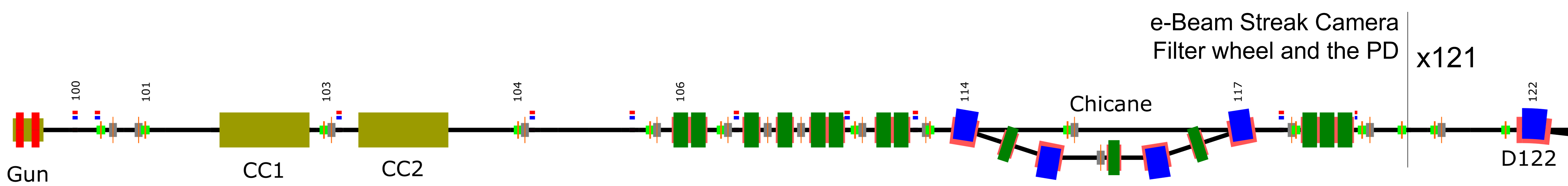}
\caption{FAST layout used in the experiment. A chicane follows two cryogenic RF cavities, compressing the bunch before diagnostics cross X121, where an Al-coated Si substrate acts as an optical transition radiation (OTR) screen. All the quadrupoles except the last triplet before the cross are turned off.}
\label{fig1}
\end{figure}

The data were collected from the existing diagnostics cross X121 (see Fig.~\ref{fig1}), where a YAG:Ce scintillator screen and an OTR screen are installed and viewed by a CCD and a streak camera respectively. The bunch parameters were thus controlled in three dimensions by a chicane and quadrupoles.

\begin{figure}[htb]
\centering 
\includegraphics[width=0.6\textwidth]{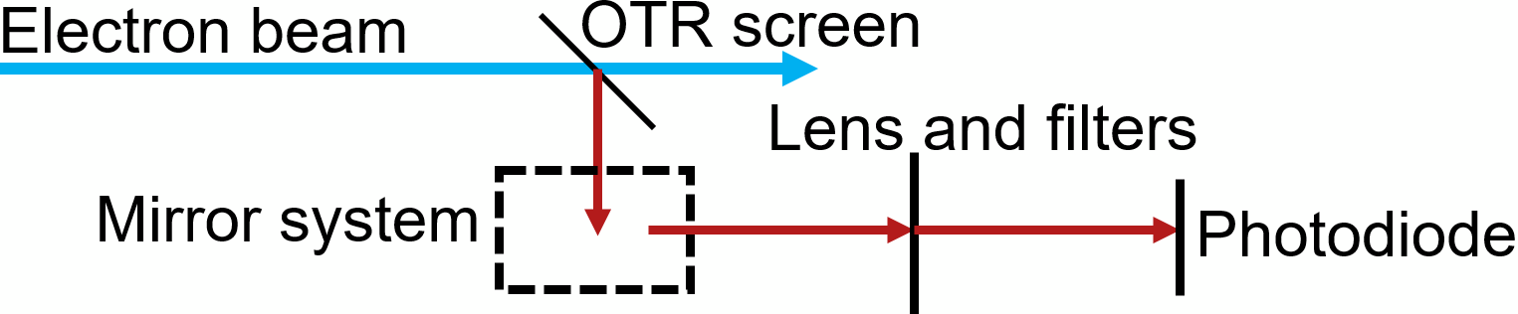}
\caption{Principle experimental scheme: TR spectrometer.}
\label{fig2}
\end{figure}

The principal experimental apparatus, shown in Fig.~\ref{fig2}, starts with an Al-coated Si substrate - a thin highly reflective foil (the OTR source), inserted into the bunch line at $45^\circ$ with respect to the bunch direction. The emitted radiation energy is transported through an optical channel, passed through one of the available 12 NIR band-pass filters (BPFs) with bandwidth centers ranging from 750 to 2400~nm, and focused onto a sensitive photodiode.

The transport line consists of two parabolic mirrors with the foil and the photodiode located in the corresponding focal points, and six highly reflective flat mirrors. The optical signal leaves the vacuum chamber through $\text{SiO}_2$ (fused silica) glass window with a cutoff transmittance wavelength of $3~\mu$m. Detailed specifications of the optical channel and detector are summarized in the appendix \ref{ap:optical_channel}.

\begin{figure}[htb]
\centering
\includegraphics[width=.48\textwidth]{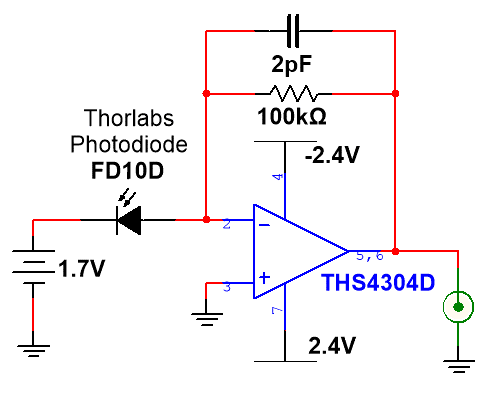}
 \quad
 \includegraphics[width=.48\textwidth]{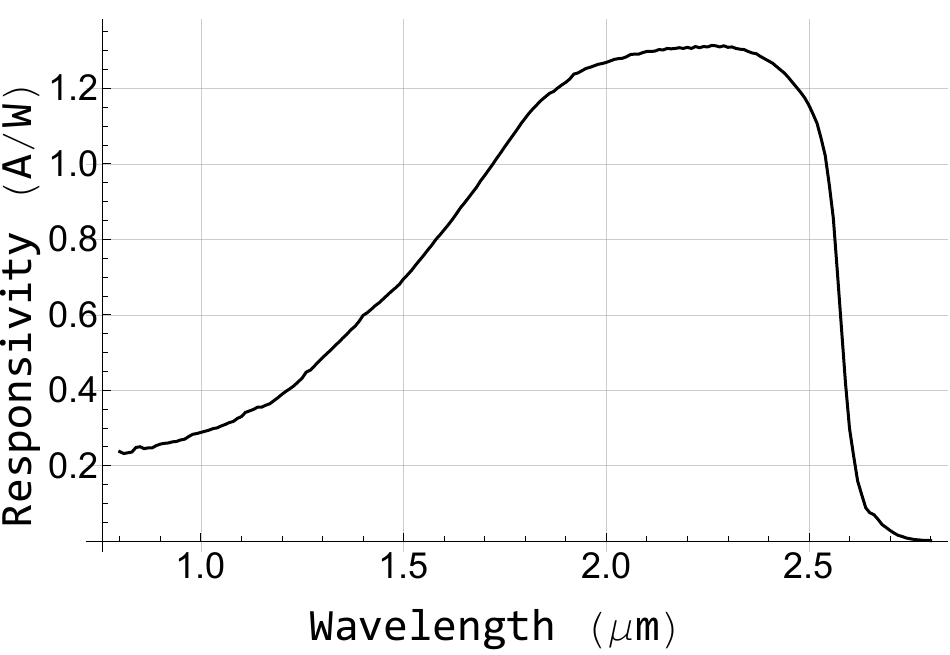}
\caption{\label{schematic} Schematic of the amplifier circuit (left) and spectral response curve of the Thorlabs InGaAs photodiode FD10D (right).}
\end{figure}

The photodiode signal is transformed to voltage by means of a high-gain integrating amplifier, shown in Fig. \ref{schematic} alongside with the diode responsivity. This detector has low response at the red end of the visible-light regime, and maximum response in the 1800-2500~nm regime. The detector was calibrated in the same way as the previous amplifier \cite{nagaitsev:napac2022-mopa34} using two different laser systems.

The transmission functions of the filters used were measured at UChicago Soft Matter Characterization Facility using the Shimadzu – UV-3600 Plus (UV-VIS-NIR Spectrophotometer) with three detectors (PMT, InGaAs, PbS), allowing accurate transmission measurements in a wide wavelength range (185-3300~nm).

\begin{figure}[htb]
\centering
\includegraphics[width=.48\textwidth]{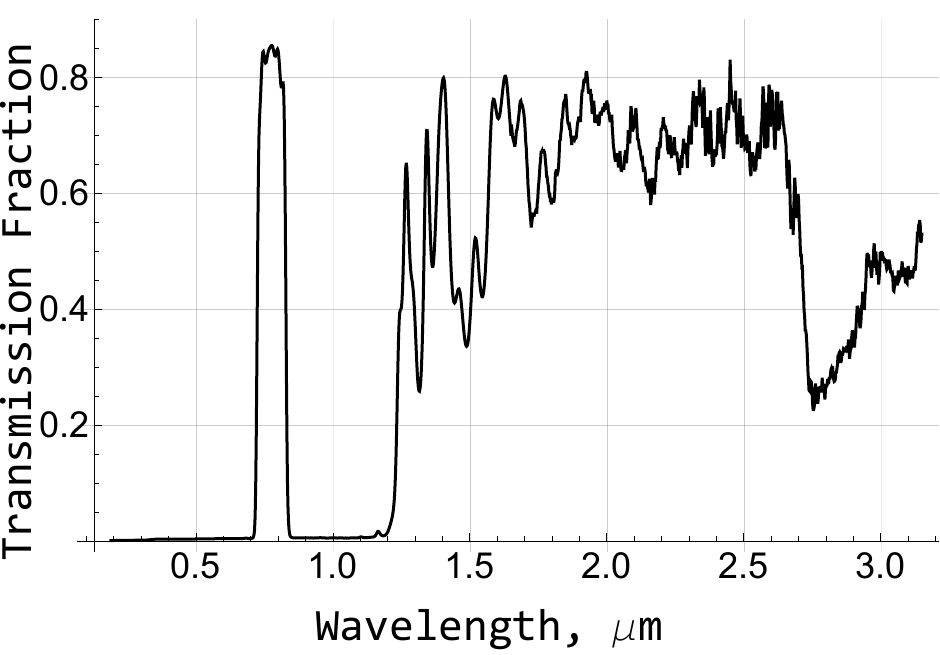}
 \quad
\includegraphics[width=.48\textwidth]{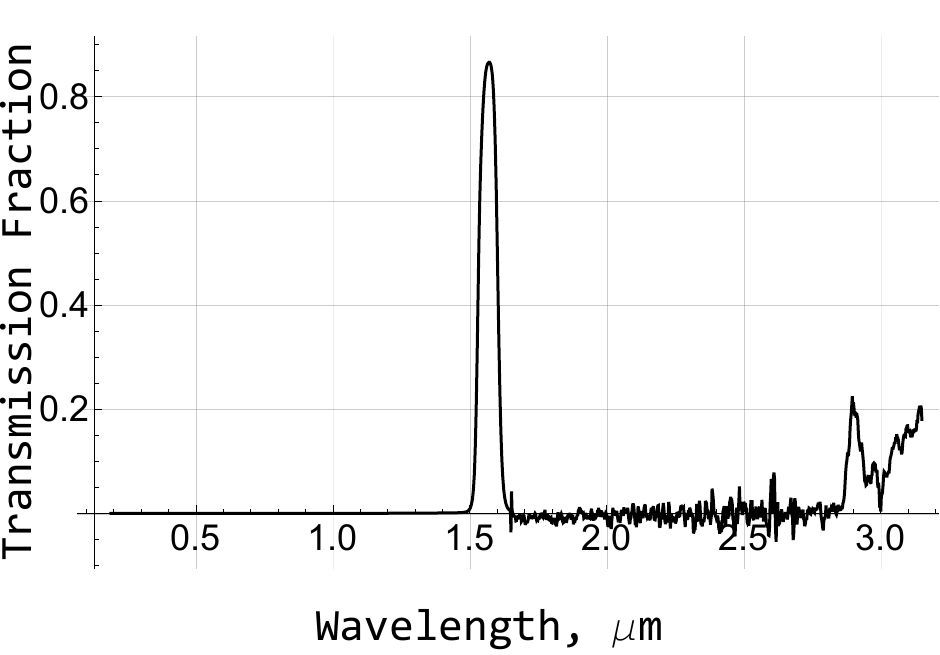}
\caption{\label{filter_transmissions} 770~nm (left) and 1570~nm (right) filter transmission functions.}
\end{figure}

Two examples of the measurement results are presented in Fig.~\ref{filter_transmissions}. 
The functions being far from the desired 100-nm-bandwidth ideals result in a significant complication to the data analysis (see section \ref{sec:theory}).

\section{Theoretical expectations}
\label{sec:theory}

\subsection{OTR spectrum}

\textbf{Single particle.} The OTR energy spectrum of one particle with charge $q$, passing through a metallic foil (screen), is well described in literature \cite{wiedemann}:

\begin{equation}
      \frac{dW_1}{d\omega d\Omega} = \frac{Z_0 q^2}{16 \pi^3} \left[\frac{\boldsymbol{n}\times\boldsymbol{\beta}}{1+\boldsymbol{\beta} \cdot \boldsymbol{n}}+\frac{\boldsymbol{n}\times\boldsymbol{\beta}}{1-\boldsymbol{\beta} \cdot \boldsymbol{n}}\right]^2
      = \frac{Z_0 q^2}{4 \pi^3} \left| \boldsymbol{n}\times \boldsymbol{\beta} \right|^2 \left[\frac{1}{1-(\boldsymbol{\beta} \cdot \boldsymbol{n})^2}\right]^2.
\end{equation}

Here $\boldsymbol{n}$ is the normal unit vector to the screen and $\boldsymbol{\beta}$ is the particle velocity vector $\boldsymbol{v}$ divided by $c$. Integration over the angles leads to the desired energy spectrum:

\begin{equation}
\frac{dW_1}{d\omega}=\int{\frac{dW_1}{d\omega d\Omega}\sin{\theta}d\theta d\phi},
\end{equation}

where the integration is taken by the half-space into which $\boldsymbol{n}$ points. It is convenient to choose the "$z$" axis along $\boldsymbol{n}$. In this case, the angle $\theta$ corresponding to the incident particle velocity $\boldsymbol{\beta}$ is determined by $\cos{\theta_0} = \frac{\boldsymbol{\beta}\boldsymbol{n}}{\beta}$. It is interesting to compute the integral for the specific case of $\theta_0=0$:

\begin{equation}
    \frac{dW_1}{d\omega}= 2 \pi \int_0^{\pi/2} \frac{d^2W_1}{d\omega d\Omega} \, \sin \theta \, d\theta =
    \frac{Z_0 q^2}{4 \pi^2} \left [ \left ( \frac{1}{\beta} + \beta \right ) \mathrm{atanh} \, \beta - 1 \right ].
    \label{otr_spectrum}
\end{equation}

In our case, the incident angle is close to $\pi/4$: $\theta_0=\pi/4+p_x$, where $p_x$ is the particle transverse momentum in radians. In practice, the integration boundaries are determined by the geometric restrictions, and the integral is computed numerically.

It will be helpful in the data analysis to express the energy spectrum in terms of the radiation wavelength:

\begin{equation}
    \frac{dW_1}{d\lambda}= \frac{dW_1}{d\omega}\, \frac{2 \pi c}{\lambda^2}.
\end{equation}

\textbf{N particles.} With omitted common multiplier, if we ignore the transverse bunch sizes at the foil, the radiated energy spectrum can be written using the Fourier integral components \cite{LANDAU1984394} of electric fields:

\begin{equation}
\label{n_part_main_eq}
    \frac{dW}{d \omega}\propto E_{\omega} E_\omega^* \propto\sum_{n=1}^N{e^{i \omega t_n}} \sum_{m=1}^N{e^{-i\omega t_m}}=N+\sum_{\begin{smallmatrix}
  m,n\\
  m\neq n
\end{smallmatrix}}{\mathrm{exp}[i \omega (t_n-t_m)]},
\end{equation}

where \(t_n\) is the arrival time of the $n^{\text{th}}$ particle ($n=1 ... N$). Here we have assumed that the bunch is ultra-relativistic, and all the radiation is directed forward. The first and second terms are called "incoherent" and "coherent" transition radiation respectively. If a bunch has transverse size larger than the observed wavelength, an additional factor $p(w)$ appears before the second term. Note that the expression can be rewritten as

\begin{equation}
    \frac{dW}{d \omega}=\frac{dW_1}{d\omega}N^2\left|\rho(\omega)\right|^2.
\label{actualOTRandSpectrumConnection}
\end{equation}

For instance, $N=1$ results in the single-particle spectrum, and a quiet bunch with $N$ particles gives

\begin{equation}
  \left|\rho(\omega)\right|^2=\frac{1}{N} \qquad \text{and} \qquad \frac{dW}{d \omega}=N \frac{dW_1}{d\omega}.
\end{equation}

It is convenient to relate the OTR spectrum to the spectrum of a quiet bunch using \textbf{Fano factor}. Usually, it is used for electrons in conductors, and characterizes the spectrum of these electrons arriving to a cross-section of the conductor \cite{PhysRev.72.26}. A beam of electrons in an accelerator arriving to an OTR screen is not any different from this conventional case.

The Fano factor can be obtained directly from the definition \cite{Beenakker_quantumShotNoise}:

\begin{equation}
    F(\omega) = \frac{S(\omega)}{2q\overline{I}} = \frac{1}{2I(\omega=0)\Delta \omega}\int_{\omega-\frac{\Delta \omega}{2}}^{\omega+\frac{\Delta \omega}{2}}{\left[\left|\delta I(-\omega)\right|^2+\left|\delta I(\omega)\right|^2\right]d\omega}\approx \frac{1}{I(\omega=0)}{\left|\delta I(\omega)\right|^2},
\end{equation}

where $\delta I(\omega) = I(\omega) - I_0(\omega)$ is the current modulation, $I_0(\omega)$ is the unperturbed current without shot noise, and the approximation is made for the case when the current modulations are smooth. It is convenient to use the distribution function, that is normalized by 1, instead of the current: $I(\omega) = \sum_{j=1}^N{\text{e}^{-i\omega t_j}} = N\rho(\omega)$. Here $t_j$ is the time when a $j$-particle reaches the OTR screen. The smooth unperturbed distribution function $\rho_0(\omega)$ is an exponential zero at the frequencies we are interested in, meaning that $\delta\rho(\omega)$ can be substituted with $\rho(\omega)$. With these changes, the Fano factor and the diffusion rate from Eq. \ref{DdifftoDcool1} can be expressed in convenient forms:

\begin{equation}
\begin{array}{c}
     F(\omega)=N\left|\delta \rho(\omega)\right|^2, \\ [0.2cm]
     \frac{T_\text{cool}}{T_{\text{diff}_e}}=r_2 \frac{\int_{-\infty}^{\infty}{\left|Z_{e,2}(\omega)\right|^2 F(\omega) d\omega}}{\int_{-\infty}^{\infty}{\left|Z_{e,2}(\omega)\right|^2 d\omega}}.
\end{array}
\end{equation}

\subsection{Total power}

It is convenient to use wavelength notations $\lambda$ instead of frequency $\omega$ when propagating the OTR spectrum through the measurement system. The voltage \(U = \frac{Q}{C} = \frac{1}{C}\int{I_c(t) dt}\) on the capacitor due to the accumulated charge is

\begin{equation}
    U = \frac{1}{C}\int{I_c(t) dt} = \frac{1}{C}\int{r(\lambda) P(t,\lambda) dt d\lambda} = \frac{1}{C}\int{r(\lambda) E(\lambda) d\lambda},
\end{equation}

where $C$ is the capacitance, $I_c$ is the current generated by the photodiode, $r(\lambda)$ is the PD responsivity in \(\left[\frac{\text{A}}{\text{W}}\right]\), presented in Fig.~\ref{schematic}, and $E(\lambda)$ is the spectrum of the radiation energy passed through a filter. Here we assumed that all the charge, produced by the PD, settles on the capacitor because the bunch length is much shorter than the amplifier circuit RC time. This equation establishes a reversible connection of the energy at the PD with the measured voltage.

On the other hand, the map \(\frac{dW}{d \omega} \rightarrow E(\lambda)\), connecting the TR spectrum before the transport channel and total energy passed through a filter is not reversible because of the limited number of filters and their wide transmission functions. Therefore, we choose the direct map at all three steps of the transformation from the bunch distribution to the set of observed signals (voltages) \(\rho(\lambda)\rightarrow \{U_i\}\), and fit the data by repeatedly taking guesses of the distribution:


\begin{equation} U_i = \int{T_i(\lambda) \frac{N \left|\rho(\lambda)\right|^2}{\lambda^2} d\lambda} = \int{T_i(\lambda) F(\lambda)\frac{1}{\lambda^2} d\lambda},\end{equation}

where the Fano factor is \(F(\lambda) = N \left|\rho(\lambda)\right|^2\) and $T(\lambda)$ is the transmission function of the transport line (optical channel), including the photodiode responsivity and the capacitance $C$.

\subsection{Optical channel}
\label{sec:lightchannel}

Detailed derivation with explanations of the geometric restrictions of the optical channel can be found in appendix \ref{ap:optical_channel}. The calculations result shows that position of a ray at PD does not depend on the emission angle, and is determined solely by the particle position in the bunch. Moreover, the phase advance between rays, emitted into different direction, is zero.

It means that the spectral field intensity at a given point on the PD cannot depend on the bunch transverse size, and is determined by the Fourier transform of the bunch longitudinal distribution function at a particular transverse point on the OTR screen. The transverse size dependence appears when we integrate by this initial point on the screen.

The calculations can be done again for the CMOS camera, which is placed further away from the second parabolic mirror than the focal distance. In this case, of course, the OTR angular distribution is imaged.

the photodiode aperture $r_{PD}$ gives a very restricting boundaries for the allowed particle positions $x_0$ at the OTR screen:

\begin{equation}\left|x_0\right|_{\text{max}} = \frac{r_{PD}}{\sqrt{2}} = 0.035~\text{cm}.\end{equation}

The mirrors and filter aperture restrictions depend almost only on the emission angle $\theta$ and give

\begin{equation}\left|\theta\right|_{\text{max}}=0.074.\end{equation}

The main source of the signal loss is the PD aperture: with the bunch size of \(\sigma_\perp=0.6\)~mm it leads to a coefficient of 0.15. Note that this factor is applied to the whole signal.

\subsection{Transverse bunch size dependence}

The transverse bunch size reduces the coherence of the signal, leaving the incoherent signal intact. The effect in a simple system is widely studied \cite{orlandi2006transvSizeEffects,chiadroni2012TransvSizeEffects}. However, the results obtained in these studies are not directly applicable here due to a different measurement system. Indeed, the exponent employed there, that effectively damps the transition radiation power to zero on wavelengths smaller than the bunch size, was obtained by integration by the phase advance between different radiation rays (directions). Our experimental setup was designed to suppress the coherent transition radiation damping by removing this relative phase advance (see appendix \ref{ap:optical_channel}), so the coherence is weakened only due to the finite width of the radiation components.

Let us start the explanation from the extreme case, when the transverse bunch size is large enough, so that the minimal transverse distance between a pair of adjacent particles is much larger than the wavelength of interest. In this case, the electromagnetic fields do not interfere, and the total power is the incoherent power. Now, if all rays, emitted from a point on the OTR screen, land onto a single point of the photodiode plane, and the phase does not depend on the emission angle and the point position, the fields do interfere in this point. They also interfere with the rays landing nearby, that are not further than one ray width away, that is taken to be equal to one wavelength. The rays displaced further apart do not interact with each other, similarly to the extreme case described above. Therefore, we estimate the transverse size impact factor $p$ by dividing the bunch into not interfering with each other cylinders with transverse size $s=\pi\, \lambda^2$:

\begin{equation}
  p = \frac{dW}{d\omega}\left(\frac{dI_{0}}{d\omega}\right)^{-1} = \frac{Q_c^2}{Q^2} n_c = \frac{1}{N^2} \left(\frac{N}{n_c}\right)^2\;n_c = \frac{\lambda^2}{\sigma_\perp^2}.
\end{equation}

Here $I_0$ is the radiation power with assumed perfect transverse coherence, $N$ is the total number of particles in the bunch, $Q_c$ is the mean charge of a cylinder, and $n_c$ is the number of the cylinders. With $\lambda = 1$~$\mu$m and $\sigma_\perp=1$~mm, an estimation of the transverse coherence factor is $p\approx10^{-6}$.

\section{Experimental results}
\label{sec:exp_results}

\begin{figure}[htb]
\centering 
\includegraphics[width=.48\textwidth]{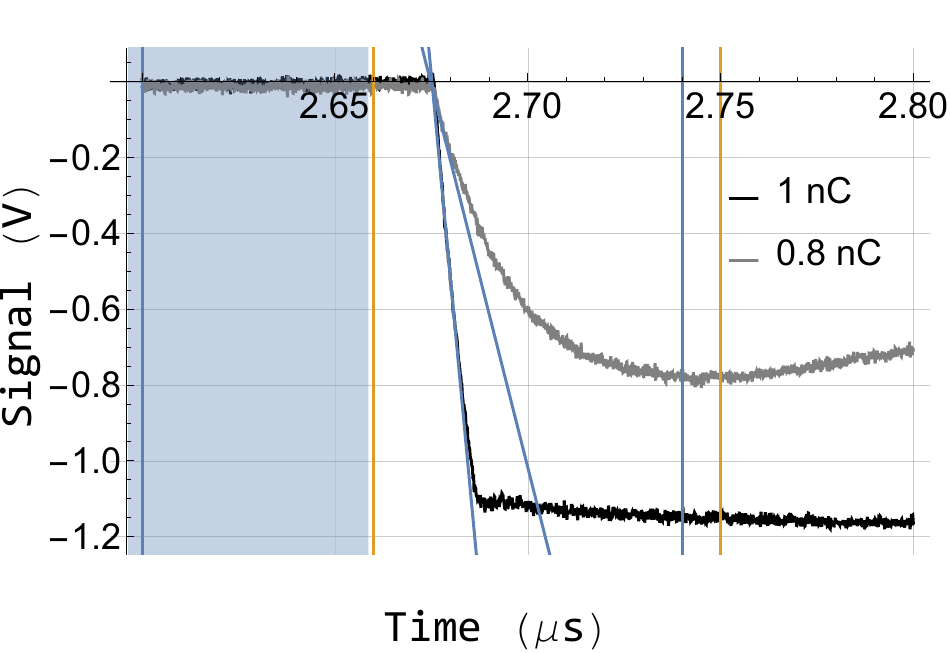}
\caption{Amplifier response to a single-bunch transition radiation pulse for two different bunch charges. The bunch here is compressed, the signal is amplified due to the coherent optical transition radiation (COTR). If the signal saturates, its amplitude is evaluated by its rising edge, approximated as a line. If not, the amplitude is found by subtraction of the dark current, averaged by the dashed area, from the signal extremum, averaged by the area between the second pair of vertical lines. For example, the signal for 1~nC is evaluated here as $2.24\pm0.08$~V, and for 0.8~nC as $0.7\pm0.01$~V.}
\label{TR_signal}
\end{figure}

The amplifier response to the TR signal is a typical curve of an RC circuit, where the RC time constant $\tau=230$~ns is much larger than the bunch length $l = 10$~ps and than the photodiode rise time $t_p=25$~ns. In most measurements the signal is averaged by 20 single-shots. The signal saturation is treated by applying a fit to the rising part, which slope is proportional to the signal amplitude. An example of an overshot signal is presented in Fig~\ref{TR_signal}.

\subsection{Low-level NIR noise}

\begin{figure}[htb]
\centering
\includegraphics[width=.48\textwidth]{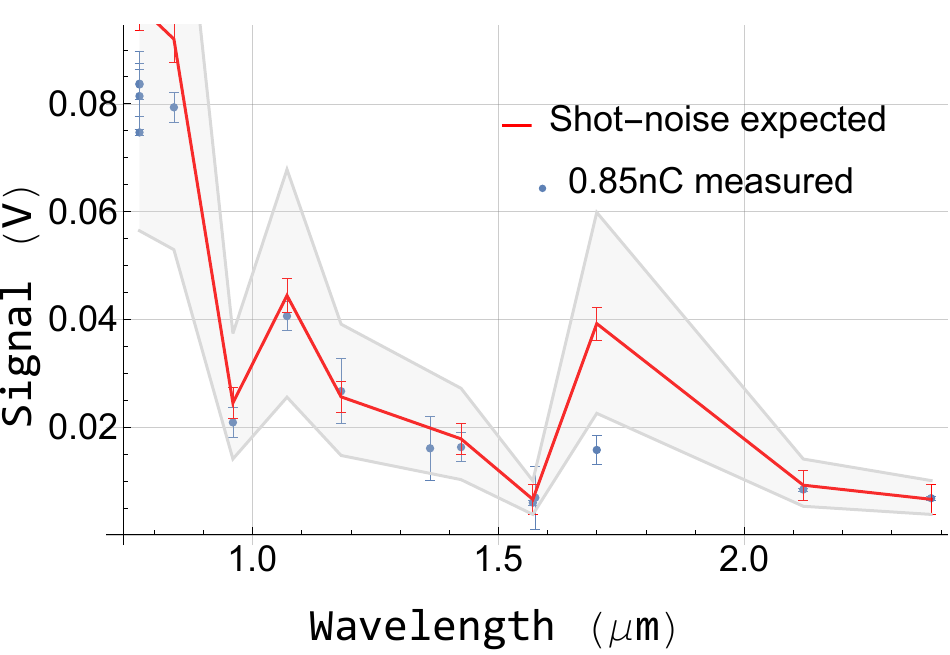}
 \quad
\includegraphics[width=.48\textwidth]{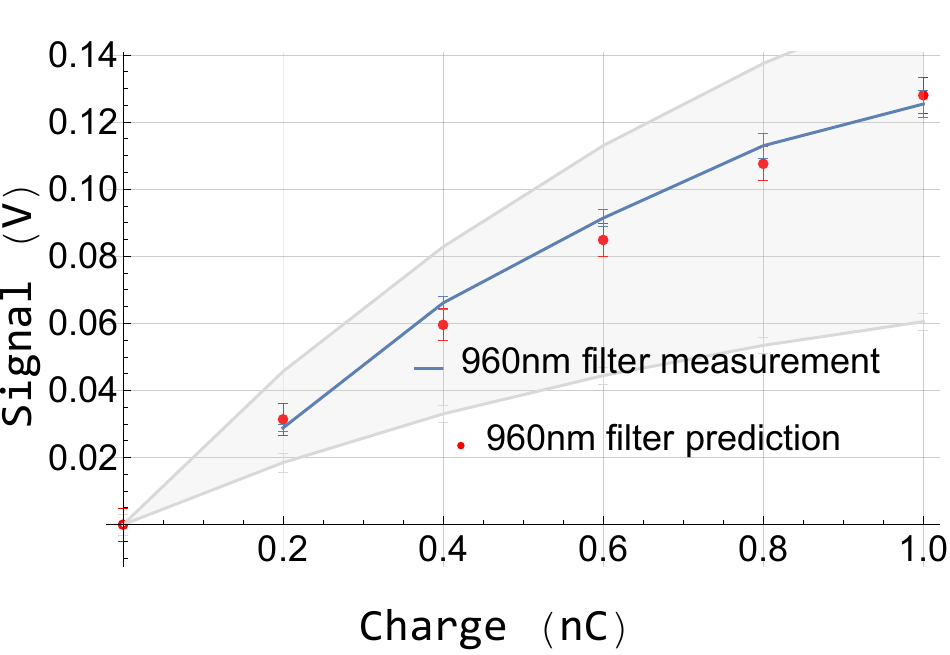}
\caption{\label{uncompressed} Experimentally measured voltage induced by the optical transition radiation (OTR) and its comparison with the predictions based on the shot-noise bunch model. Left: comparison at different filters; right: comparison for different charges. The shaded area is a systematic uncertainty estimation.}
\end{figure}

Comparison of the theoretical predictions for a shot-noise beam and experimental results is presented in Fig.~\ref{uncompressed}. Here the combined reflectivity of the optical channel is taken to be $80\pm10$~\%, and the transverse bunch shape - round Gaussian. X-axis values are the means of the band-pass filters specified by the manufacturer; red is a set of the predicted voltage values, induced by the radiation passed through a set of filters; blue values are the measured voltages.

The uncertainties have two components:

\begin{itemize}
    \item Individual uncertainties at each filter: uncertainties in the scope data, uncertainties of the filter transmission function measurements, bunch charge fluctuations;
    \item Common multiplier at each filter: systematic uncertainties;
\end{itemize}

Individual uncertainties are depicted at both experimental and theoretical curves as error bars, while the systematic uncertainties are shown as bands.

The common multiplier uncertainty is composed from the uncertainties in the detector and beam distribution measurement accuracy: flat mirrors and vacuum window transmission functions, elements misalignment, bunch size measurement uncertainty, and finally the uncertainty of the calculations because of the bunch shape not being round. In total, it becomes $\Delta F=30\%$.

The random uncertainty, composed of the bunch size and charge fluctuations, uncertainties of the filter transmission functions measurements, and signal fluctuations on the scope, are much smaller: $\Delta F=5\%$.

As a result, the Fano factor for the EIC CEC bunch parameters at 25 MeV can be estimated as $F=1\pm\frac{0.3}{p}$, where $p=0.08\pm0.04$ is the transverse coherence factor (see sec. \ref{sec:comparison}). Of course, the Factor cannot be negative, so instead of writing answers in $\pm$ form, we use a one-sided confidence interval. Therefore, with 67\% confidence

\begin{equation}
\begin{array}{c}
    F < 5,  \\[0.2cm]
    \frac{T_\text{cool}}{T_{\text{diff}_e}} < 0.1. 
\end{array}
\end{equation}

\subsection{Elevated NIR noise}
\label{sec:elevated_noise}

The OTR signal increase has been previously observed in experiments with external modulation presence \cite{lumpkinCompressionInFEL,SASEfelMicrobunching}, which gives a peaked gain (\(\frac{\text{COTR}}{\text{OTR}}\)), and in experiments without any external excitation \cite{LoosLCLSCOTR,weschFLASHOTR}, giving a gain that is monotonically increasing with wavelength, starting at around 200~$\mu$m. Usually, the elevation becomes visible with bunch compression, when the characteristic length of the wavelength spectrum moves into the detector sensitivity range.

The FAST lattice used in this research is different from the facilities mentioned above in energy: $\gamma=$~30-50 instead of $\gamma=$~200-400: the bunch is not rigid, and the particles are not tightly bound longitudinally. Therefore, we expect the elevation to appear at even smaller wavelengths.

\begin{figure}[htb]
\centering
\includegraphics[width=.48\textwidth]{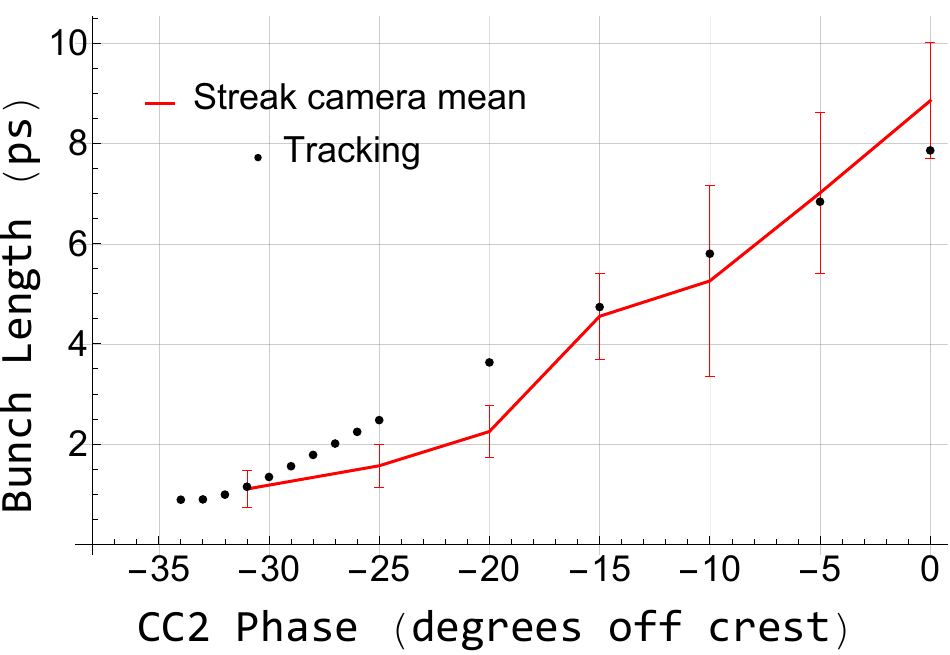}
 \quad
\includegraphics[width=.48\textwidth]{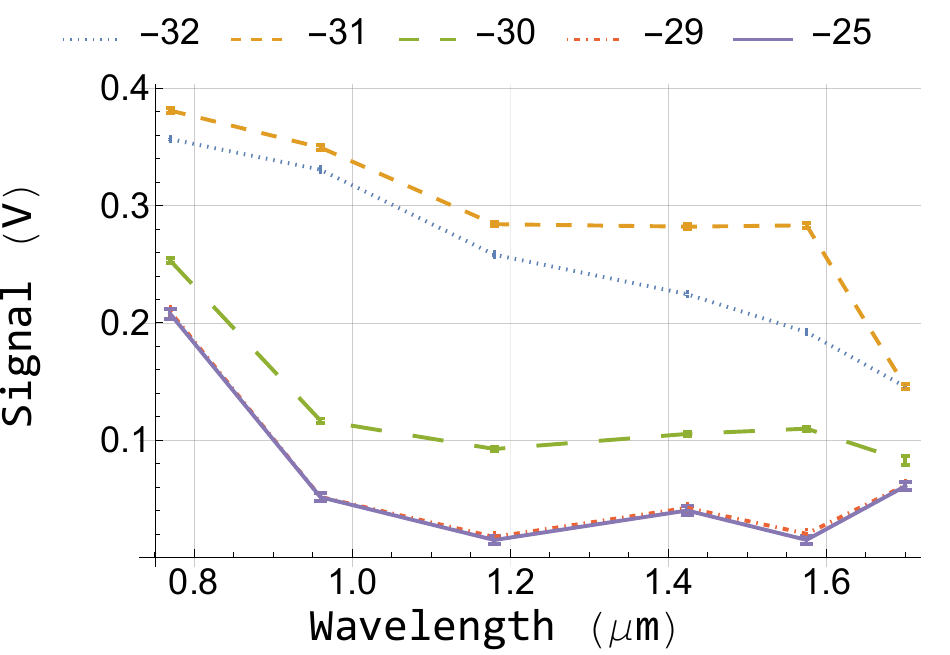}
\caption{Left: bunch length dependence on the second RF cavity (CC2) phase $\phi$, obtained from the streak camera measurements and tracking simulations; the on-crest value is $\phi=0^\circ$. Right: the elevated signal for various $\phi$, listed above the figure, and filters, arranged by the X-axis.}
\label{beamlength}
\end{figure}

Bunch length dependence on the second RF cavity (CC2) phase $\phi$ is presented in Fig.~\ref{beamlength} left, alongside with the simulated curve (see section \ref{sec:simulations}). The maximum compression factor achieved is $c=8$ (from 8.5~ps to 1~ps), with the energy of the bunch kept constant.

Uncompressed and compressed bunch signals comparison is presented in Fig.~\ref{beamlength} right. Here the bunch parameters are adjusted so that the signal increase is small for visibility. Although the increase in the signal is observed at all filters, it should be noted that the gain in emitted OTR cannot be obtained directly from the data because the filters have high leakage.

Taking broadband spectra as fit guesses gives unexpectedly poor results, predicting much higher gain at low-wavelength filters than it is observed experimentally, especially at 770~nm. It is due to the high broad leakage these filters possess at higher wavelengths. On the contrary, filters with 1570, 2120 and 2380~nm labels are "pure", meaning that their transmission curves are almost like specified by the manufacturer.

\begin{figure}[htb]
\centering
\includegraphics[width=.48\textwidth]{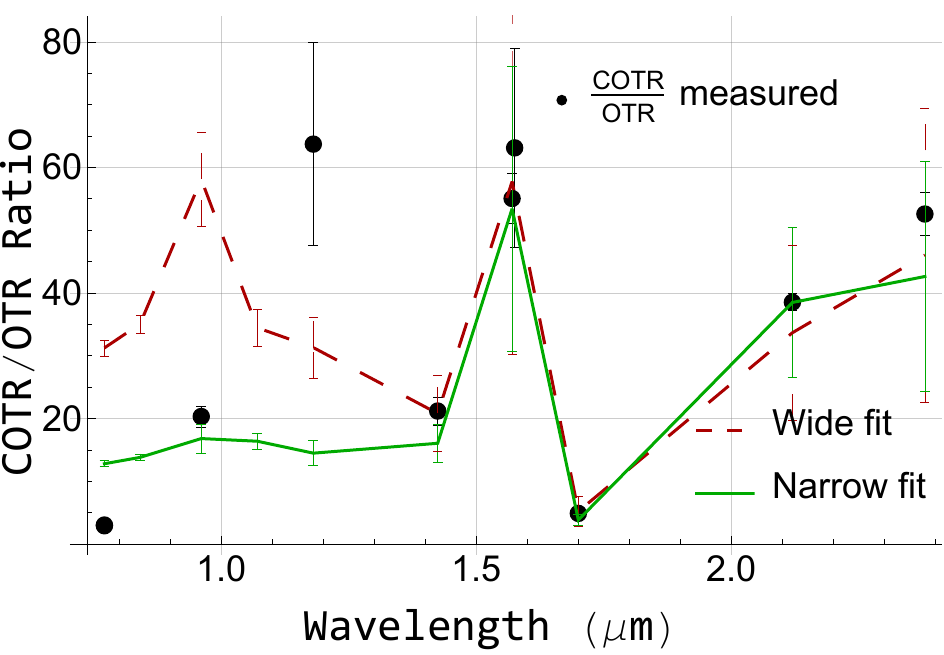}
 \quad
\includegraphics[width=.48\textwidth]{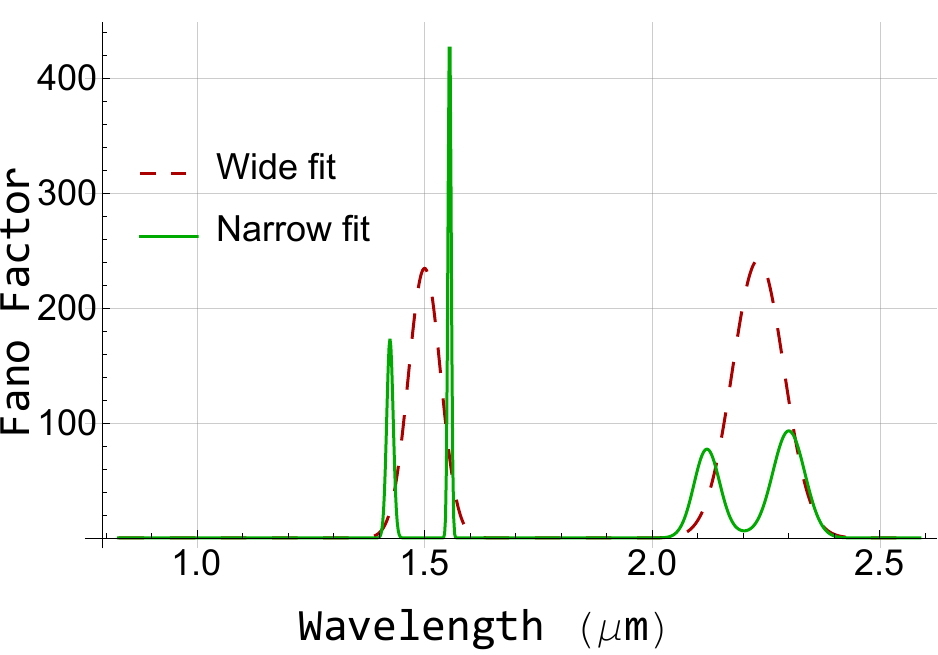}
\caption{Left: signal predictions based on a guessed spectrum and their comparison with the experimental data. Right: the guessed spectrum presented in terms of the Fano factor.}
\label{compressed}
\end{figure}

Based on these features, it can be concluded that bunch spectrum has a spike at around $1.5~\mu$m, and a broad elevation at wavelengths $\lambda > 2~\mu$m. Two examples of possible fits are presented in Fig.~\ref{compressed}.

The maximum observed signal increase is around 50. If the transverse coherence factor was $p=10^{-6}$, the actual microbunching would be of the order of $10^{-2}$ from the bunch density maximum. Of course, the $p=10^{-6}$ is only an estimation based on the assumption that the coherence length is only one wavelength long.

\begin{figure}[htb]
\centering
\includegraphics[width=.48\textwidth]{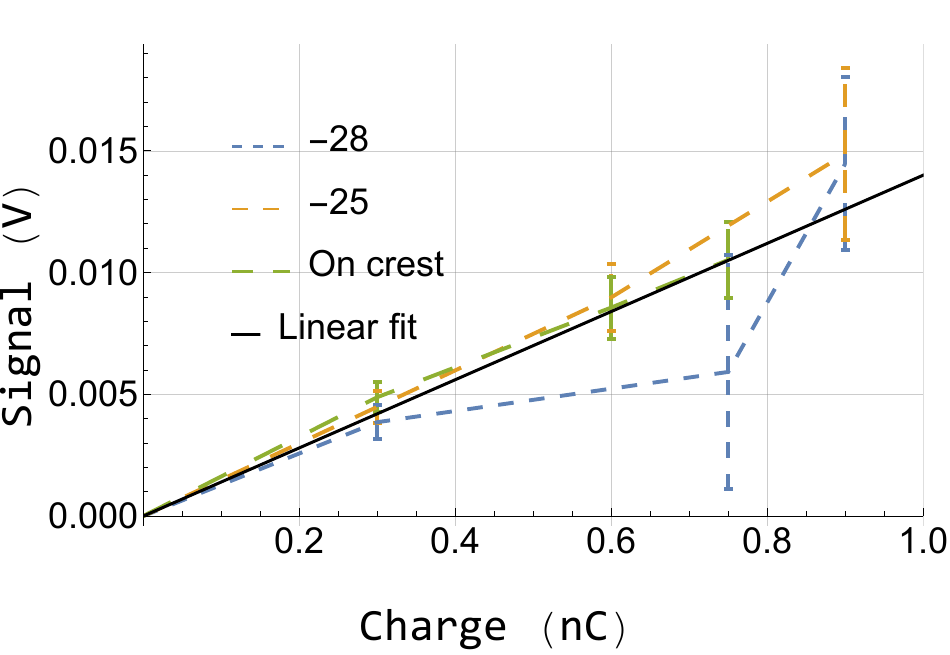}
 \quad
\includegraphics[width=.48\textwidth]{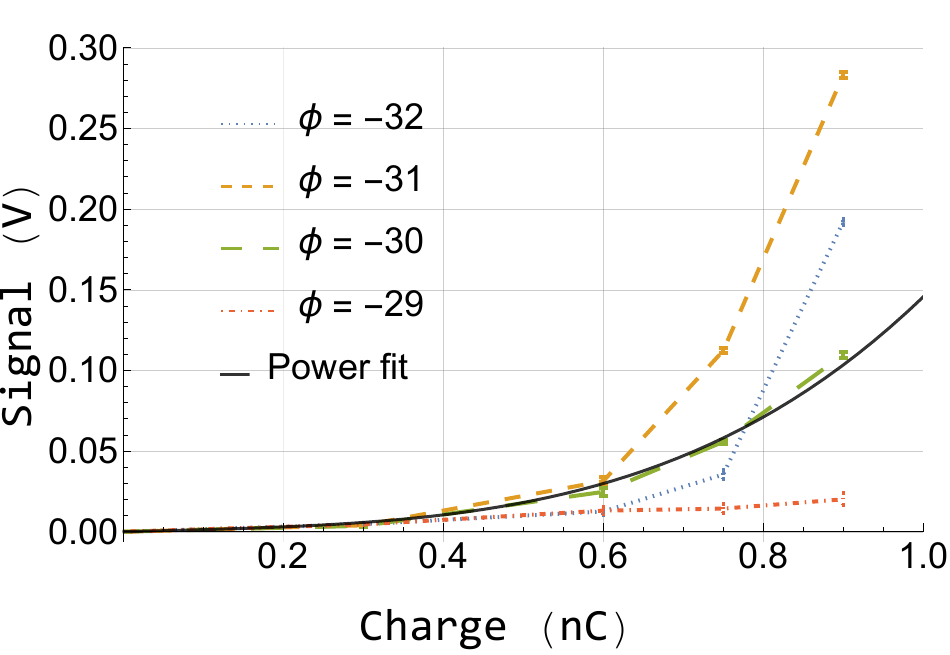}
\caption{Charge dependence of the observed signal for uncompressed (left) and compressed (right) bunch for different second RF cavity (CC2) phases $\phi$, listed on the plots in degrees. Black line on the right is a power law fit with the power of $\frac{7}{2}$. The power is different from the expected $Q^2$ mostly because the bunch length, and hence the positions and intensity of the OTR spectrum peaks, depends on the bunch charge.}
\label{charge_dep}
\end{figure}

Signal dependence on the bunch charge for various acceleration cavity phases is presented in Fig.~\ref{charge_dep}. The shot-level signal (uncompressed bunch, left side) shows a linear dependence, while compressing the bunch (right side) leads to a highly nonlinear behavior.

\begin{figure}[htb]
\centering
\includegraphics[width=.48\textwidth]{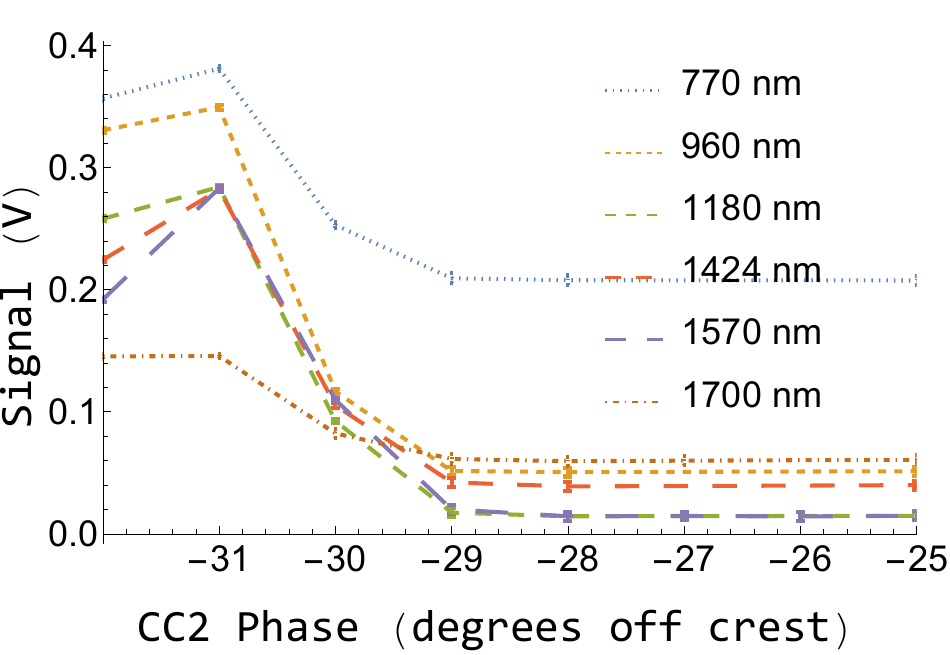}
 \quad
\includegraphics[width=.48\textwidth]{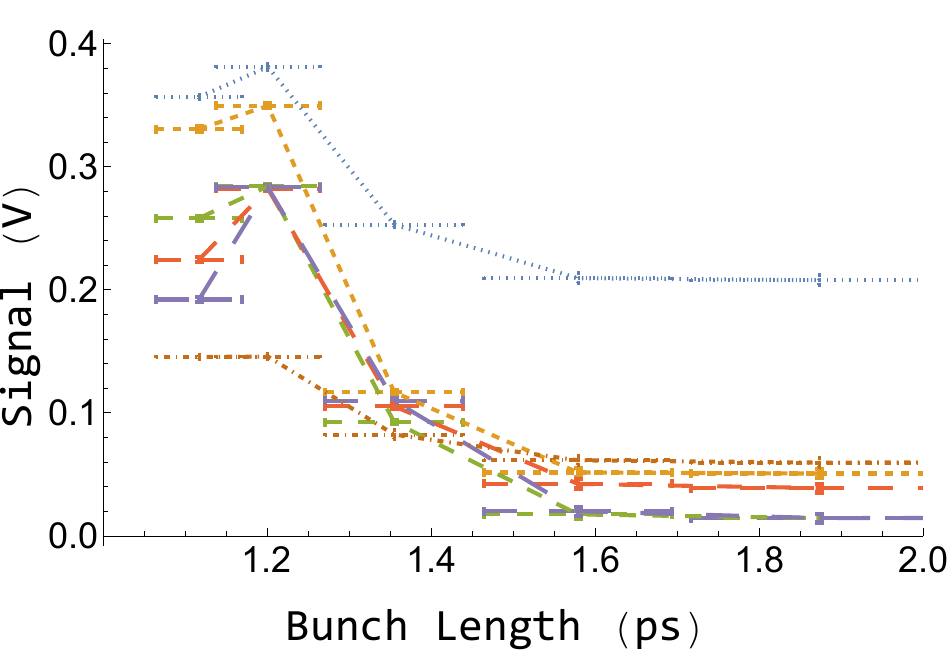}
\caption{\label{OTRcc2Length} Dependence of the OTR power on the second RF cavity (CC2) phase $\phi$ (left), and on the bunch length $\sigma_z$ (right). The bunch length is calculated from the simulations results.}
\end{figure}

The OTR power dependence on the second RF cavity (CC2) phase $\phi$ and bunch length $\sigma_z$ is shown in Fig.~\ref{OTRcc2Length}. It is interesting that the difference in the extrema positions of the OTR power and bunch length dependence on $\phi$ can be used as a way to determine the initial amplitude of the energy modulations. In the first approximation it can be written as

\begin{equation}
    \Delta E \approx E_{\text{cav}} \frac{\omega_c}{\omega}\left(\sin{\phi_1}-\sin{\phi_2}\right),
\end{equation}

where $E_{\text{cav}}$ is the energy that a particle gains from the second RF cavity (CC2) when it is on-crest, $\omega_c$ is the cavity frequency, $\omega=\frac{2\pi c}{\lambda}$ is the bunch distribution modulation frequency, $\phi_1$ is the CC2 phase where the bunch length is minimal, and $\phi_2$ is the CC2 phase where the OTR radiation spectral component at wavelength $\lambda$ is maximal. Of course, the high leakage of the filters used here makes this method too inaccurate, but getting a rough estimation is still possible.

\textbf{OTR power fluctuations}

The data were fitted assuming that the incoherent power did not change after the bunch compression. This assumption can be easily verified experimentally by looking at the intensity fluctuations. Indeed, if the histograms of the compressed and uncompressed bunch OTR signals intersect, according to the eq. \ref{n_part_main_eq}, it means that the incoherent level was reduced. However, we don't see any intersection on the experiment.

\begin{figure}[htb]
\centering
\includegraphics[width=.48\textwidth]{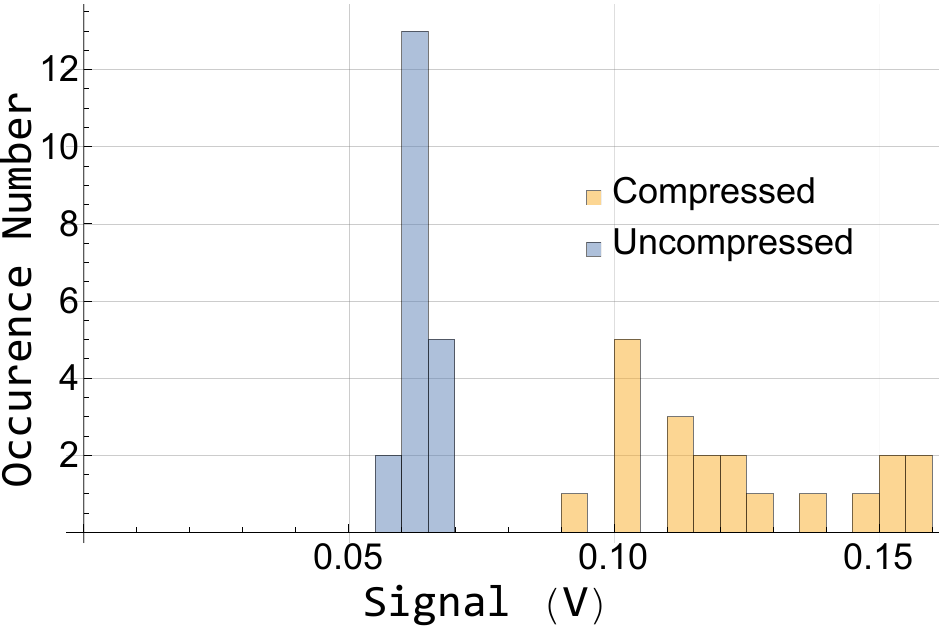}
 \quad
\includegraphics[width=.48\textwidth]{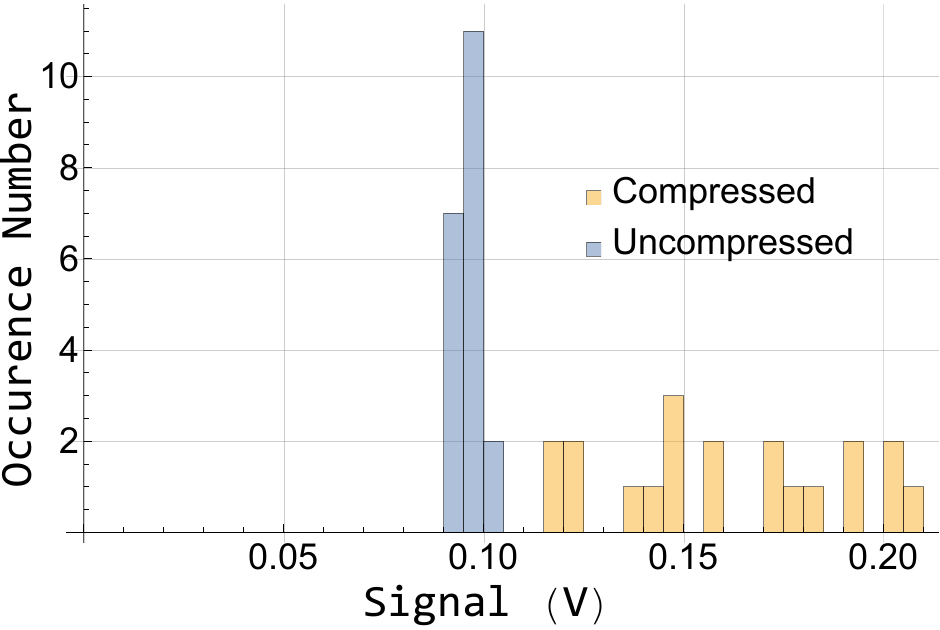}
\caption{\label{hists} Signal amplitude fluctuations  for 1424 and 1700~nm filters at 1~nC. The distribution width is much larger if the signal is enhanced (compressed bunch), suggesting a coherent process starting from noise.}
\end{figure}

The OTR intensity histograms for 1424 and 1700~$\mu$m filters at 1~nC for uncompressed (no microbunching) and compressed (with microbunching) bunches are presented in Fig.~\ref{hists}. The large fluctuations of the signal for compressed bunch are consistent with a coherent process starting from noise; the longitudinal-space-charge-induced microbunching in this case.

\section{Simulations}
\label{sec:simulations}

The space charge tracking was performed using ASTRA \cite{ASTRA} at the cathode and accelerating part of the FAST lattice, followed by an s-based code ImpactX \cite{impactX} responsible for the chicane. The space charge is taken into account using 3D Poisson solvers: FFT algorithm in ASTRA and "multigrid" in ImpactX. All the simulations are run on a personal computer, with ImpactX being parallelized on a GPU.

\begin{figure}[htbp]
\centering
\includegraphics[width=.47\textwidth]{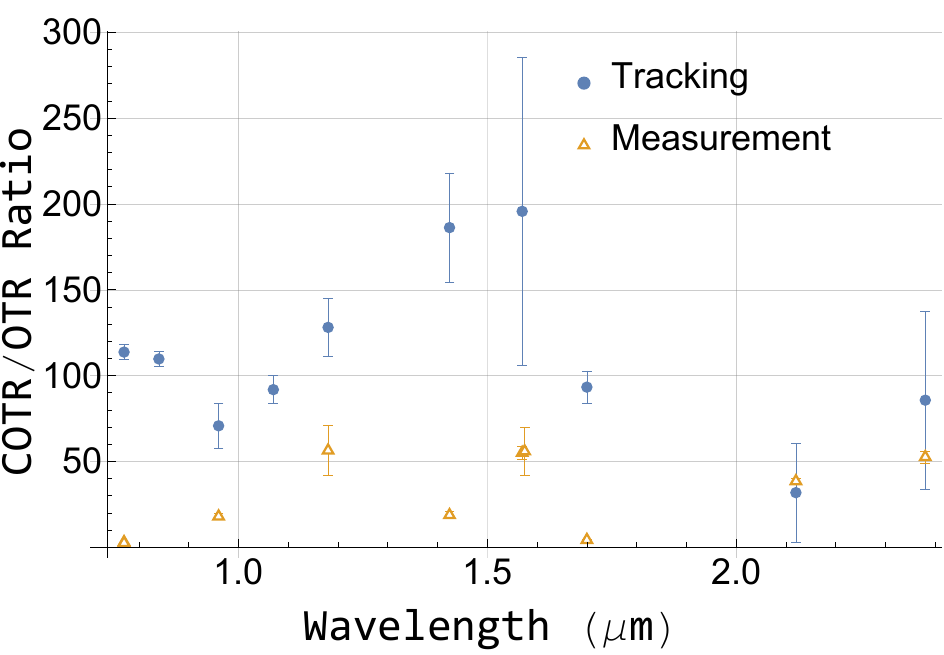}
\qquad
\includegraphics[width=.47\textwidth]{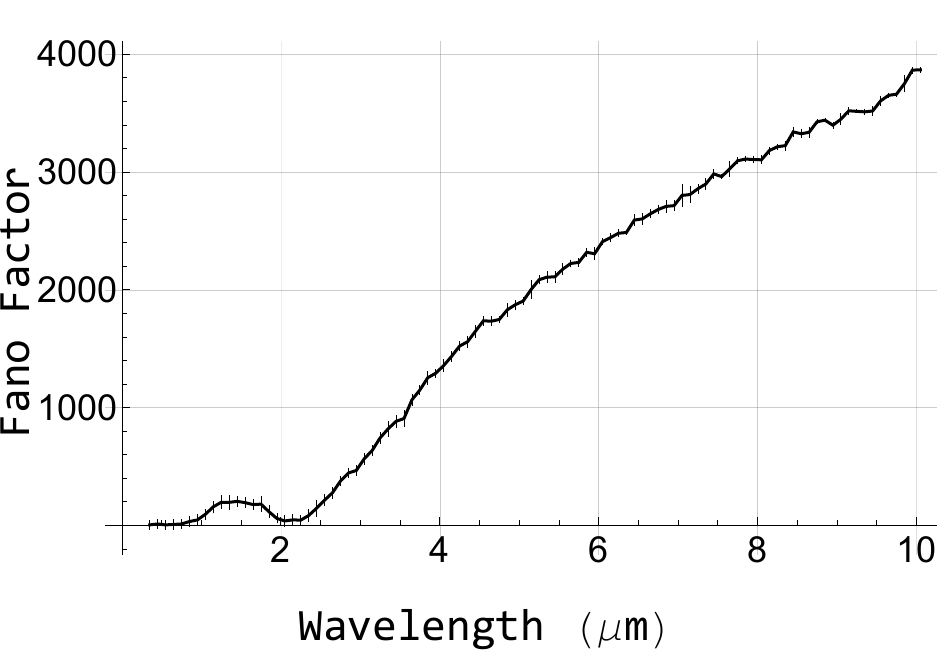}
\caption{Left: comparison of the raw experimental data, presented as the ratio of the signals from compressed and uncompressed beams, and the same values but calculated from simulation results, with an assumption of the transverse coherence factor being $p=1$. Total least squares fit gives $p=0.08\pm0.04$. Right: the corresponding spectrum of the longitudinal bunch distribution, presented in terms of the Fano factor.}
\label{fig:simulatedFanoAndRatios}
\end{figure}

The simulations give an agreement with the experiment in terms of general bunch properties, as it can be seen, for example, from Fig.~\ref{beamlength}. We can transform the spectra into the expected signal using the same procedure as for the guessed spectra in section \ref{sec:elevated_noise}. The data points obtained for each filter are presented in Fig.~\ref{fig:simulatedFanoAndRatios} left, alongside with the experimental values for the second RF cavity (CC2) phase $\phi=-31^\circ$ and the bunch charge is $Q=1$~nC. The final simulated bunch longitudinal distribution spectrum in units of the Fano factor is presented in Fig. \ref{fig:simulatedFanoAndRatios} right.

\begin{figure}[htbp]
\centering
\includegraphics[width=.47\textwidth]{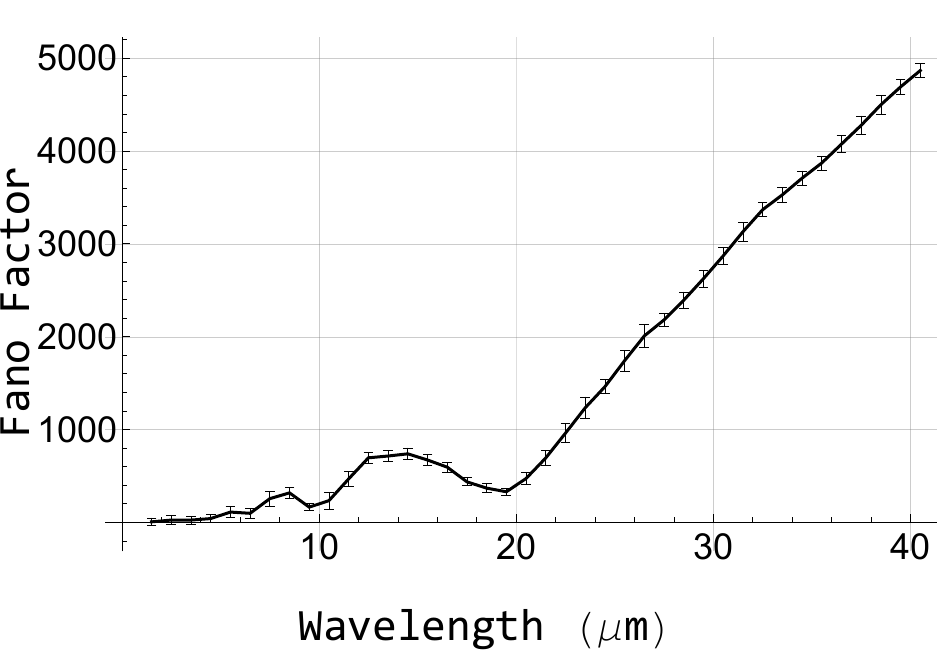}
\qquad
\includegraphics[width=.47\textwidth]{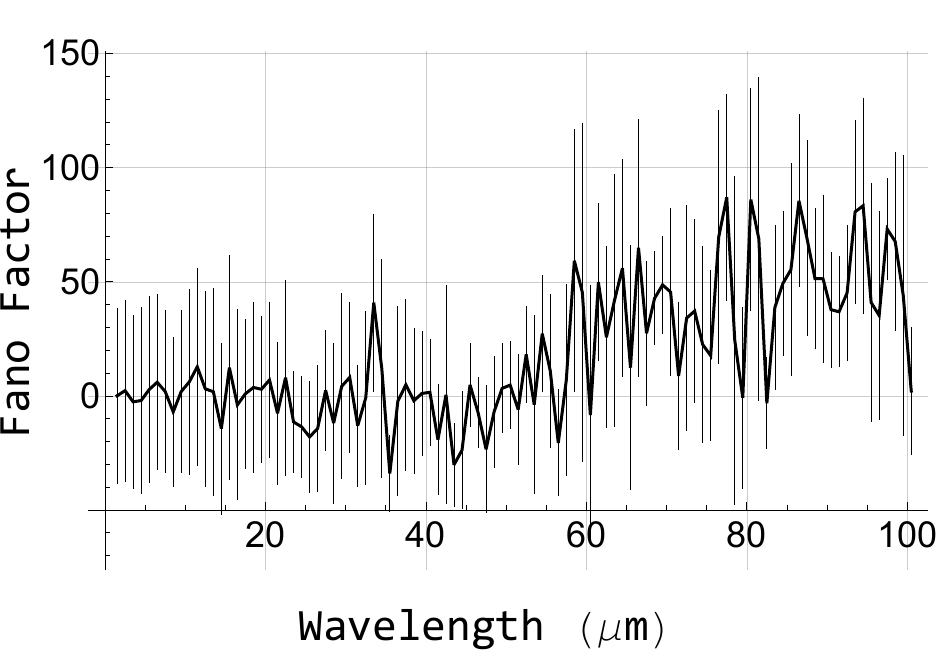}
\caption{Same as Fig.~\ref{fig:simulatedFanoAndRatios} right. Fano factor obtained from tracking for the second RF cavity (CC2) phase $\phi=-20$ (left) and $\phi=0$ (right).}
\label{fig:moreSimFanoFactors}
\end{figure}

The simulations present a valuable distinction from the experiment: the spectrum is known everywhere. Therefore, by establishing a connection between two approaches in the intersecting region ($0.7-2~\mu$m) we can estimate the experimental results at any wavelength for any compression factor. For instance, the Fano factors obtained through tracking are presented in Fig.~\ref{fig:moreSimFanoFactors} for the second RF cavity (CC2) phases $\phi=-20$ (left) and $\phi=0$ (right). A constant fit of the Fano factor for the uncompressed bunch ($\phi=0$) is $F=0\pm10$ all the way up to $\lambda=50~\mu$m - a range that covers the entire area of the EIC CEC impedance significance (see Fig.~\ref{CECimpedance}). The result was obtained from a list of fits with various bunch distribution binning and spectrum averaging. Again, the Fano factor $F$ cannot be less than 0, so we introduce an upper bound: with 67\% confidence the Fano factor is less than 10:

\begin{equation}
\begin{array}{c}
    F < 10,  \\[0.2cm]
    \frac{T_\text{cool}}{T_{\text{diff}_e}} < 0.2. 
\end{array}
\end{equation}

\section{Comparison of the results}
\label{sec:comparison}

The uncompressed or moderately compressed bunch (\(l>2\)~ps) has no additional noise above the shot level \(F=1\), which is supported by both the experiment and the space-charge particle tracking. This baseline is conserved after the bunch compression, with the coherent part revealing itself and reaching maximum at the second RF cavity (CC2) phase $\phi=-31^\circ$. By comparing the simulated and experimental spectra for compressed bunches, we derive $p\approx0.08\pm0.04$, which deviates significantly from our estimation (by a factor of $10^5$) and and is vastly different (by a factor of $\sim10^{150}$) from the calculations used in experiments with small transverse beam sizes. The factor smallness is a good point for a further study of how exactly the radiation is formed, transported in a double-parabolic-mirror channel, and collected by a photodiode.

However, the individual filters' data are not in agreement. The difference is most likely caused by the systematic uncertainties of the experiment: the transmission functions of the elements are not known exactly, and the bunch parameters were not controlled with enough precision due to a lack of theoretical predictions and tracking at the time of the experiment.

Nevertheless, other possible issues were also considered. The bunch energy is too high for wakefields to have such a major impact. Moreover, the characteristic short-range wakefield wavelength is $5$~mm $\gg1$~$\mu$m, and the lattice length is too short. Moving further, the bunch current is not high enough for CSR to noticeably change the bunch dynamics in the NIR region.

The signal increase might be explained by another type of radiation, reflected from the OTR screen. Examples include COSR or Coherent Optical Edge Radiation (COER). The radiation traveling directly from the bending magnets to the diagnostics system, as well as any other similar direct radiation, was ruled out by blocking the path. Still, it can couple to the beam pipe waveguide modes, land on the OTR screen, and be reflected to the optical channel. For the estimation purposes it can be assumed that all the radiated energy couples to the waveguide. With this assumption, two radiation candidates have the same power dependence on the longitudinal bunch distribution of the bunch as the COTR (here we also assume that the last dipole is the only source of the radiation, and the bunch length change is small both in the dipole and between the dipole and the OTR source). All is left to do is to compare the relative amplitude of these three.

Simple estimates, based on \cite{wiedemann} give us the following ratio of CTR:CSR:

\begin{equation}
    \frac{dW_{TR}}{d\omega} : \frac{dW_{SR}}{d\omega} = \frac{e^2}{4\pi \varepsilon_0 c}\left( \frac{1}{\pi^2} \int_0^\pi{\frac{\beta^2 \sin^3{\theta} \; d\theta}{\left(1-\beta^2\cos^2{\theta}\right)^2}} : \frac{\sqrt{3}\gamma \omega}{2\omega_c} \int_{\omega/\omega_c}^\infty{\text{K}_{5/3}}(x) \; dx \right),
\end{equation}

where $\omega_c = \frac{3}{2} \omega_0 \gamma^3$ is the critical frequency, $\omega_0 = \frac{\beta c}{r}$ is the angular frequency of the particles moving in a bending magnet, $r$ is the bend radius, and $\text{K}$ is the modified Bessel function. The ratio is 64:1 at $\lambda=3$~$\mu$m, with CSR becoming negligible if the wavelength is even lower.

Generally, the CER is noticeable only if the wavelength is larger than the critical wavelength of the synchrotron radiation, which is 28~$\mu$m in our case. Furthermore, in order to be reflected properly (so that to eventually land on the photodiode), the radiation has to be inside a narrow area on the OTR screen, as discussed in section \ref{sec:lightchannel}. Because a general waveguide mode is not tightly constrained within this area, we need to introduce an additional factor of \(\frac{r_{\text{PD}}^2}{2r_p^2}\approx2\times10^{-4}\) for any radiation generated not on the OTR screen itself. Next, we need to get rid of the assumption about perfect collection of the synchrotron radiation from the whole fourth dipole, and the constant minimal bunch length along the curvature. Last, but not least, the radiation reflected from the OTR screen is subject to the same transverse bunch size factor $p$, where the "bunch size" should be substituted with the OTR screen radius, that makes the factor even smaller. These factors together, without the transverse bunch size factor, total to a value less than \(10^{-5}\), effectively damping the radiation amplitude observed on the photodiode, meaning that the signal increase is not due to CSR or CER.

As a result, we state that the signal elevation is indeed caused by a microbunching structure in the bunch, that creates the coherent term in transition radiation spectrum. The transverse coherence factor is then taken to be $p=0.08\pm0.04$, not dependent on wavelength.

\section{Conclusion}

We have constructed and analyzed a low-level noise measurement system, collected the experimental data, and compared it with theoretical expectations and simulations. An uncompressed or moderately compressed bunch (\(l>2\)~ps) has no additional noise above shot level \(F=1\), that is supported by both experiment and space-charge particle tracking. The experiment covers the range of wavelengths up to $\lambda=2.5~\mu$m, while the rest of the EIC CEC significance range is covered by the simulations.

With 67\% confidence, the electron diffusion rate is less than 10\% of the cooling rate ($\frac{T_\text{cool}}{T_{\text{diff}_e}} < 0.1$) according to the experiment uncertainties, and less than 20\% of the cooling rate ($\frac{T_\text{cool}}{T_{\text{diff}_e}} < 0.2$) according to the simulations uncertainties. Noticeable microbunching starts to appear at $\lambda=60~\mu$m. For comparison, the same ratio for the hadron bunch diffusion is $\frac{T_\text{cool}}{T_{\text{diff}_h}}=0.04$.

A compressed bunch (\(l\approx1\)~ps), on the contrary, shows large microbunching amplitudes, especially on wavelengths $\lambda>2.5$~$\mu$m, as can be seen from the simulations. The modulations were confirmed experimentally, though only in the spectral segment with smaller increases, as the detector sensitivity extends only up to 3~$\mu$m. Extending this comparison to a broader spectral range in future studies could improve the accuracy of the noise level measurement.

An estimation of the transverse bunch size effect on coherence loss in an optical channel was proposed to better align with experimental observations. While this approach demonstrated significantly improved accuracy compared to the small bunch approximation, the values obtained from comparison with simulations showed some disagreement.

Future studies could expand the experimental spectral range at higher energies to better align with the requirements of CEC schemes and FELs. These experiments should have precise diagnostics of all the measurement elements and bunch parameters, as well as include a theoretical investigation of coherence conservation effects to enhance the accuracy of the measurements. The measurement system could benefit from using better band pass filters that don't have a leakage, together with neutral density filters to avoid the signal saturation.

\acknowledgments

This project is supported by a grant (DE-SC0022196) from the DOE ARDAP office.  This work was also supported by the U.S. National Science Foundation under Award PHY-1549132, the Center for Bright Beams, and by the University of Chicago.

This manuscript has been authored by Fermi Research Alliance, LLC under Contract No. DE-AC02-07CH11359 and by Brookhaven Science Associates, LLC under Contract No. DE-SC0012704 with the U.S. Department of Energy, Office of Science.

This research used the open-source particle-in-cell code ImpactX https://github.com/ECP-WarpX/impactx. We acknowledge all ImpactX contributors.

We are grateful to the UChicago Soft Matter Characterization Facility staff and to Philip Griffin in particular for allowing us to use their spectrophotometer in our needs.

\newpage

\appendix

\section{Electron bunch noise effects in the EIC CEC system}
\label{ap:noise_and_EIC_CEC}

\renewcommand{\theequation}{A.\arabic{equation}}
\setcounter{equation}{0}

The cooling times with noise effects have been well described in \cite{stupakov2019moreNoiseInvestigations} for the noise passing through the amplification sections (and acting in the kicker), and also in \cite{stupakov2018coolingShotNoiseEffect} for the modulator. Different effects, including the interference between the noise passed through the modulator and the shot noise at the beginning of the amplifiers, are summarized in \cite{bergan2024electronNoiseInCEC}.

We first use Eqs. 39-40 from \cite{stupakov2019moreNoiseInvestigations} with the EIC CEC parameters from \cite{eicTechReport} to verify that the parameters used are the same, obtaining 165 minutes of the cooling time. Next, we calculate the $r_2$ factor from Eqs. 48, 67-68 of the same article to get

\begin{equation}
    r_2=\frac{T_\text{cool}}{T_{\text{diff}_e}}\vert_\text{shot-noise}=0.021.
\end{equation}

Now we compare the diffusion rates with and without modulator effects using the impedances provided in Eq.~14 of \cite{bergan2024electronNoiseInCEC}. In this derivation we follow the notations from \cite{bergan2024electronNoiseInCEC}, including usage of the wave vector $k = \frac{\omega}{c}$ and coordinate $z = c t$. It is more convenient to not go to the space domain and utilize the impedances: we transform Eq.~16 from the same article to get a change of the hadron bunch energy in one turn:

\begin{equation}
    \Delta \eta_{h,i}(z)=\frac{r_h}{\gamma}\int{ Z_i(k)\delta \rho_i (k)\text{e}^{i k z} dk},
    \label{Dnotsquared}
\end{equation}

where index $i$ corresponds to the terms from Eq.~16 of \cite{bergan2024electronNoiseInCEC}.

By averaging the absolute value squared of \ref{Dnotsquared} by $z$ and particles $i$, we arrive to the same answers as in \cite{bergan2024electronNoiseInCEC}, but for an arbitrary noise. For example, for the noise in electron bunch, presented at the entrance to the amplifier, we have:

\begin{equation}
\label{eq:diffusion}
    \langle \Delta \eta_{h}^2 \rangle_{e,2} = \left(\frac{r_h}{\gamma}\right)^2 n_e \int{\left|Z_{e,2}(k)\right|^2\left|\delta \rho_e(k)\right|^2 dk}.
\end{equation}

Note that for the shot-noise bunch the result is the same as in \cite{bergan2024electronNoiseInCEC} because of the Parseval's theorem. The diffusion rate is defined as

\begin{equation}
    \frac{1}{T_\text{diff}} = \frac{1}{\eta_{h}^2}\;\frac{\mathrm{d}\eta_{h}^2}{\mathrm{d}t} = 
    \frac{1}{\eta_{h}^2}\;\frac{\langle \Delta \eta_{h}^2 \rangle}{T},
\end{equation}

where $T$ is the turn period. Last, we use \ref{eq:diffusion} to finally derive the ratio of two diffusion rates: caused by the noise at the modulator (1) and by the electron shot noise at the entrance to the amplifier (2):

\begin{equation}
    \frac{T_{\text{diff}_{e,2}}}{T_{\text{diff}_{e,1}}}=\frac{\int_{-\infty}^{\infty}{\left|Z_{e,1}(k)\right|^2}dk}{\int_{-\infty}^{\infty}{\left|Z_{e,2}(k)\right|^2}dk} \approx 0.7\times10^{-3}
\end{equation}

Therefore, we can safely neglect the effect in the modulator and the interference term, and obtain Eq. (\ref{DdifftoDcool1}), with $k$ instead of $\omega$. Of course, transformation to frequencies requires only switching $k$ to $\omega$.

\begin{equation}
    \frac{T_\text{cool}}{T_{\text{diff}_e}}=r_2 \frac{\int_{-\infty}^{\infty}{\left|Z_{e,2}(k)\right|^2\left|\delta \rho_e(k)\right|^2 dk}}{\int_{-\infty}^{\infty}{\left|Z_{e,2}(k)\right|^2 \frac{1}{N} dk}}.
\end{equation}

Following the explanation in the introduction,  $\left|\delta\rho_e(\omega)\right|^2$ can be substituted with  $\left|\rho_e(\omega)\right|^2$. We can get the bunch spectrum $\left|\rho_e(\omega)\right|^2$ from Eq. (\ref{actualOTRandSpectrumConnection}) and express the electron diffusion-to-cooling ratio in terms of the measurements:

\begin{equation}
    \frac{T_\text{cool}}{T_{\text{diff}_e}}=r_2 \frac{\int_{-\infty}^{\infty}{\left|Z_{e,2}(\omega)\right|^2}\frac{dW}{d\omega} d\omega}{\int_{-\infty}^{\infty}{\left|Z_{e,2}(\omega)\right|^2 N \frac{dW_1}{d\omega} d\omega}},
\end{equation}

where $ \frac{dW_1}{d\omega}$ is the OTR spectrum from one particle, that does not depend on $\omega$, and $\frac{dW}{d\omega}$ is the OTR spectrum measured experimentally.

Parameters used in the text are presented in table \ref{tab:EICCECparams}.

\begin{table}[!hbt]
   \caption{\label{tab:EICCECparams}%
   Main parameters of the Electron Ion Collider Coherent Electron Cooling (EIC CEC) system.}
   \begin{ruledtabular}
   \begin{tabular}{lr}
       \textbf{Parameter} & \textbf{Value}\\
       \colrule
           Modulator length $L_m$, m &  40 \\ 
           Kicker length $L_k$, m &  40 \\ 
           Hadron particle charge $Z$, e &  1 \\ 
           Bunch size in the amplifier $\Sigma$ (h and e), mm &  0.67 \\ 
           Electron bunch energy $E$, MeV & 149.8  \\ 
           Electron chicane strength $R_{56,e}$, mm & 7.1 \\ 
           Hadron chicane strength $R_{56,h}$, mm & 1.95 \\ 
           Electron bunch charge $Q_e$, nC & 1 \\ 
           Hadron bunch charge $Q_h$, nC & 11 \\ 
           Electron bunch length $\sigma_{z,e}$, mm & 4 \\ 
           Hadron bunch length $\sigma_{z,h}$, mm & 60 \\ 
           Amplifier to modulator bunch size ratio $r=\frac{\Sigma}{\Sigma_p}$ & 0.2 \\ 
           Dimensionless length $l=\frac{\pi r}{2}$ & 1 \\ 
           Dimensionless chicane strength $q=\frac{R_{56}\sigma_\eta \gamma}{\Sigma}$ & 0.6 \\
   \end{tabular}
   \end{ruledtabular}
\end{table}

In the derivation and in the table we used that the wavelength of the oscillations is much lower than the hadron bunch size in the modulator $k\gg \frac{\gamma}{\sigma_p}$, so that the dimensionless length $l$ can be taken to be constant.

\section{Signal propagation through the optical channel}
\label{ap:optical_channel}

\renewcommand{\theequation}{B.\arabic{equation}}
\setcounter{equation}{0}

We work in the accelerator coordinates: z-axis is in the direction of the on-symmetry-axis (i.e., for the radiation transmitted from the first parabolic mirror focus point into \(\phi=0\), \(\theta=0\) direction), x-axis is horizontal, and y-axis is vertical. We assume that all elements are centered on the symmetry axis and perfectly alligned, with the OTR source and photodiode located in the focal points of the first and second parabolic mirrors respectively. The parabolic mirror equation is \(y = \frac{1}{4F} x^2 = \frac{1}{2l} x^2\), where $F$ is the focal distance of the mirror, and $l$ is the distance between the OTR source and the mirror.

By applying geometry, we get the ray positions at each mirror ($x_1, x_2$), at the filter ($x_3$), and at the photodiode ($x_4$). For analytical expressions we write the formulas in the first order by initial position $x_0$ and emission angle $\theta$:

\begin{equation}x_1 \approx l \tan{\theta} + \sqrt{2} x_0 \left(1-\sin{\theta}\right)\end{equation}
\begin{equation}\theta_1 \approx \theta - \frac{x_1}{l} \approx -\frac{\sqrt{2}x_0}{l}\end{equation}
\begin{equation}x_2 \approx \sqrt{2} x_0 + l \theta - \frac{\sqrt{2}x_0 L}{l}\end{equation}
\begin{equation}\begin{aligned}\theta_2 \approx \theta_1 + \frac{x_2}{l} \approx -\frac{\sqrt{2}x_0}{l} + \theta + \frac{\sqrt{2}x_0}{l} - \frac{\sqrt{2}x_0 L}{l^2} = \theta - \frac{\sqrt{2}x_0 L}{l^2}\end{aligned}\end{equation}
\begin{equation}x_3 \approx \sqrt{2} x_0 + d_f \left(\theta-\frac{\sqrt{2}x_0 L}{l^2}\right)\end{equation}
\begin{equation}x_4 \approx x_2 - (l+x_2)\tan{\theta_2} \approx \sqrt{2}x_0\end{equation}

The PD position does not depend on the emission angle, and is determined by the particle position in the bunch. Comparison of the PD ($r_{\text{PD}}$) and bunch ($\sigma_\perp$) sizes gives very restricting boundaries for the bunch particle positions: \(\left|x_0\right|_{\text{max}} = r_{\text{PD}}/\sqrt{2} = 0.35\)~mm. Therefore, allowed $x_0$, $y_0$ are much smaller than the mirrors size (\(x_0,y_0\ll r_{\text{PM}}\)), and can be neglected in the expressions for the coordinates on the mirrors when we compare them with apertures. As a result, the mirrors (and filter) aperture restrictions depend only on the emission angle:

\begin{equation}
\begin{array}
l\;\theta < r_{\text{PM}} \\
d_f\;\theta < r_f
\end{array}
\end{equation}

where $r_f$ is the filter radius. The filter aperture appears more restrictive and gives

\begin{equation}\left|\theta\right|_{\text{max}}=0.074\end{equation}

The system with two parabolic mirrors, employed here, generally possesses a useful feature, that the path length of the rays, emitted from the vicinity of the focal point of the first parabolic mirror, does not depend on the emission angle. The exact lengthy expression is not presented here, but it means that not only the beam shape is imaged from the radiation source onto the photodiode, but also the whole electromagnetic field (if the geometric restrictions are not taken into account).

Misalignment of the photodiode and the first parabolic mirror position relative to the OTR source lead to additional dispersion: around 20~$\mu$m phase difference per 1~mm misalignment. The parabolic mirror positioning was accurately tuned with precision higher than 0.1~mm, meaning that the misalignment effect can be dropped, leaving only the factor $p$ used in the main part of the article.

\section{Main parameters used in the article}
\label{ap:MainNEBparameters}

\renewcommand{\theequation}{C.\arabic{equation}}
\setcounter{equation}{0}

FAST specifications, important for the X121 cross simulations, are listed in table \ref{tab:fast_LEL}. More detailed information can be found at \cite{FASTlatticeGitHub}.

\begin{table}[!hbt]
   \caption{\label{tab:fast_LEL}%
   Low energy line FAST specifications.}
   \begin{ruledtabular}
   \begin{tabular}{lr}
       \textbf{Parameter} & \textbf{Value}\\
       \colrule
        RF cavities (CC1, CC2) length, m & 1.435\\ 
        RF cavities (CC1, CC2) gradient, MeV/m & 10-11 \\
        Chicane bend radius, cm & 0.8425\\ 
   \end{tabular}
   \end{ruledtabular}
\end{table}

Initial bunch (at $z=0$~m) parameters are listed in table \ref{tab:fast_beam}.

\begin{table}[!hbt]
   \caption{\label{tab:fast_beam}%
   FAST photo-cathode ($z=0$) electron bunch parameters.}
   \begin{ruledtabular}
   \begin{tabular}{lr}
       \textbf{Parameter} & \textbf{Value}\\
       \colrule
        Macro-pulse repetition rate, Hz & 1-5 \\ 
        Micro-pulse spacing, ns & 333 \\ 
        Micro-pulse length, ps & 3-20 \\ 
        bunch transverse size, mm & 0.1-1.5 \\
   \end{tabular}
   \end{ruledtabular}
\end{table}

Optical channel and detector specifications are listed in table \ref{tab:LTD}.

\begin{table}[!hbt]
   \caption{\label{tab:LTD}%
   Optical channel and detector specifications.}
   \begin{ruledtabular}
   \begin{tabular}{lr}
       \textbf{Parameter} & \textbf{Value}\\
       \colrule
        Screen to 1st mirror distance $l$, cm & 17.88\\
        1st to 2nd mirror distance $L$, cm & 415\\
        Parabolic mirrors aperture $r_{PM}$, cm & 38.1\\
        Photodiode aperture $r_{PD}$, cm & 0.05\\
        Average filter aperture $r_{f}$, cm & 0.28\\
   \end{tabular}
   \end{ruledtabular}
\end{table}

\providecommand{\noopsort}[1]{}\providecommand{\singleletter}[1]{#1}%


\begin{thebibliography}{38}%
\makeatletter
\providecommand \@ifxundefined [1]{%
 \@ifx{#1\undefined}
}%
\providecommand \@ifnum [1]{%
 \ifnum #1\expandafter \@firstoftwo
 \else \expandafter \@secondoftwo
 \fi
}%
\providecommand \@ifx [1]{%
 \ifx #1\expandafter \@firstoftwo
 \else \expandafter \@secondoftwo
 \fi
}%
\providecommand \natexlab [1]{#1}%
\providecommand \enquote  [1]{``#1''}%
\providecommand \bibnamefont  [1]{#1}%
\providecommand \bibfnamefont [1]{#1}%
\providecommand \citenamefont [1]{#1}%
\providecommand \href@noop [0]{\@secondoftwo}%
\providecommand \href [0]{\begingroup \@sanitize@url \@href}%
\providecommand \@href[1]{\@@startlink{#1}\@@href}%
\providecommand \@@href[1]{\endgroup#1\@@endlink}%
\providecommand \@sanitize@url [0]{\catcode `\\12\catcode `\$12\catcode `\&12\catcode `\#12\catcode `\^12\catcode `\_12\catcode `\%12\relax}%
\providecommand \@@startlink[1]{}%
\providecommand \@@endlink[0]{}%
\providecommand \url  [0]{\begingroup\@sanitize@url \@url }%
\providecommand \@url [1]{\endgroup\@href {#1}{\urlprefix }}%
\providecommand \urlprefix  [0]{URL }%
\providecommand \Eprint [0]{\href }%
\providecommand \doibase [0]{https://doi.org/}%
\providecommand \selectlanguage [0]{\@gobble}%
\providecommand \bibinfo  [0]{\@secondoftwo}%
\providecommand \bibfield  [0]{\@secondoftwo}%
\providecommand \translation [1]{[#1]}%
\providecommand \BibitemOpen [0]{}%
\providecommand \bibitemStop [0]{}%
\providecommand \bibitemNoStop [0]{.\EOS\space}%
\providecommand \EOS [0]{\spacefactor3000\relax}%
\providecommand \BibitemShut  [1]{\csname bibitem#1\endcsname}%
\let\auto@bib@innerbib\@empty
\bibitem [{\citenamefont {Schottky}(1918)}]{Schottky_1918}%
  \BibitemOpen
  \bibfield  {author} {\bibinfo {author} {\bibfnamefont {W.}~\bibnamefont {Schottky}},\ }\bibfield  {title} {\bibinfo {title} {Über spontane stromschwankungen in verschiedenen elektrizitätsleitern},\ }\href {https://doi.org/https://doi.org/10.1002/andp.19183622304} {\bibfield  {journal} {\bibinfo  {journal} {Annalen der Physik}\ }\textbf {\bibinfo {volume} {362}},\ \bibinfo {pages} {541} (\bibinfo {year} {1918})}\BibitemShut {NoStop}%
\bibitem [{\citenamefont {Schottky}(2018)}]{Schottky_original}%
  \BibitemOpen
  \bibfield  {author} {\bibinfo {author} {\bibfnamefont {W.}~\bibnamefont {Schottky}},\ }\bibfield  {title} {\bibinfo {title} {On spontaneous current fluctuations in various electrical conductors},\ }\href {https://doi.org/10.1117/1.jmm.17.4.041001} {\bibfield  {journal} {\bibinfo  {journal} {J. Micro/Nanolithogr. MEMS MOEMS}\ }\textbf {\bibinfo {volume} {17}},\ \bibinfo {pages} {041001} (\bibinfo {year} {2018})}\BibitemShut {NoStop}%
\bibitem [{\citenamefont {Stupakov}\ and\ \citenamefont {Baxevanis}(2019)}]{stupakov2019moreNoiseInvestigations}%
  \BibitemOpen
  \bibfield  {author} {\bibinfo {author} {\bibfnamefont {G.}~\bibnamefont {Stupakov}}\ and\ \bibinfo {author} {\bibfnamefont {P.}~\bibnamefont {Baxevanis}},\ }\bibfield  {title} {\bibinfo {title} {Microbunched electron cooling with amplification cascades},\ }\href {https://doi.org/10.1103/PhysRevAccelBeams.22.034401} {\bibfield  {journal} {\bibinfo  {journal} {Phys. Rev. Accel. Beams}\ }\textbf {\bibinfo {volume} {22}},\ \bibinfo {pages} {034401} (\bibinfo {year} {2019})}\BibitemShut {NoStop}%
\bibitem [{\citenamefont {Bergan}(2024)}]{bergan2024electronNoiseInCEC}%
  \BibitemOpen
  \bibfield  {author} {\bibinfo {author} {\bibfnamefont {W.~F.}\ \bibnamefont {Bergan}},\ }\bibfield  {title} {\bibinfo {title} {Electron diffusion in microbunched electron cooling},\ }\href {https://doi.org/10.1103/PhysRevAccelBeams.27.084402} {\bibfield  {journal} {\bibinfo  {journal} {Phys. Rev. Accel. Beams}\ }\textbf {\bibinfo {volume} {27}},\ \bibinfo {pages} {084402} (\bibinfo {year} {2024})}\BibitemShut {NoStop}%
\bibitem [{\citenamefont {Huang}\ and\ \citenamefont {Kim}(2007)}]{huang2007reviewSASEandSelfSeeded}%
  \BibitemOpen
  \bibfield  {author} {\bibinfo {author} {\bibfnamefont {Z.}~\bibnamefont {Huang}}\ and\ \bibinfo {author} {\bibfnamefont {K.-J.}\ \bibnamefont {Kim}},\ }\bibfield  {title} {\bibinfo {title} {Review of x-ray free-electron laser theory},\ }\href {https://doi.org/10.1103/PhysRevSTAB.10.034801} {\bibfield  {journal} {\bibinfo  {journal} {Phys. Rev. ST Accel. Beams}\ }\textbf {\bibinfo {volume} {10}},\ \bibinfo {pages} {034801} (\bibinfo {year} {2007})}\BibitemShut {NoStop}%
\bibitem [{\citenamefont {Saldin}\ \emph {et~al.}(2002)\citenamefont {Saldin}, \citenamefont {Schneidmiller},\ and\ \citenamefont {Yurkov}}]{saldin2002seededFELnoise}%
  \BibitemOpen
  \bibfield  {author} {\bibinfo {author} {\bibfnamefont {E.}~\bibnamefont {Saldin}}, \bibinfo {author} {\bibfnamefont {E.}~\bibnamefont {Schneidmiller}},\ and\ \bibinfo {author} {\bibfnamefont {M.}~\bibnamefont {Yurkov}},\ }\bibfield  {title} {\bibinfo {title} {Study of a noise degradation of amplification process in a multistage hghg fel},\ }\href {https://doi.org/https://doi.org/10.1016/S0030-4018(02)01091-X} {\bibfield  {journal} {\bibinfo  {journal} {Optics Communications}\ }\textbf {\bibinfo {volume} {202}},\ \bibinfo {pages} {169} (\bibinfo {year} {2002})}\BibitemShut {NoStop}%
\bibitem [{\citenamefont {Chao}\ \emph {et~al.}(2023)\citenamefont {Chao}, \citenamefont {Tigner}, \citenamefont {Weise},\ and\ \citenamefont {Zimmermann}}]{IBS_ref}%
  \BibitemOpen
  \bibfield  {author} {\bibinfo {author} {\bibfnamefont {A.~W.}\ \bibnamefont {Chao}}, \bibinfo {author} {\bibfnamefont {M.}~\bibnamefont {Tigner}}, \bibinfo {author} {\bibfnamefont {H.}~\bibnamefont {Weise}},\ and\ \bibinfo {author} {\bibfnamefont {F.}~\bibnamefont {Zimmermann}},\ }\href {https://doi.org/10.1142/13229} {\emph {\bibinfo {title} {Handbook of Accelerator Physics and Engineering}}},\ \bibinfo {edition} {3rd}\ ed.\ (\bibinfo  {publisher} {World Scientific},\ \bibinfo {year} {2023})\ pp.\ \bibinfo {pages} {192--196}\BibitemShut {NoStop}%
\bibitem [{\citenamefont {Derbenev}(2017)}]{derbenev2017ElectronCoolingTheory}%
  \BibitemOpen
  \bibfield  {author} {\bibinfo {author} {\bibfnamefont {Y.~S.}\ \bibnamefont {Derbenev}},\ }\bibfield  {title} {\bibinfo {title} {Theory of electron cooling},\ }\href@noop {} {\bibfield  {journal} {\bibinfo  {journal} {arXiv preprint arXiv:1703.09735}\ } (\bibinfo {year} {2017})}\BibitemShut {NoStop}%
\bibitem [{\citenamefont {van~der Meer}(1972)}]{vanderMeer:1972sf}%
  \BibitemOpen
  \bibfield  {author} {\bibinfo {author} {\bibfnamefont {S.}~\bibnamefont {van~der Meer}},\ }\bibfield  {title} {\bibinfo {title} {{Stochastic damping of betatron oscillations in the ISR}},\ }\href@noop {} {\bibfield  {journal} {\bibinfo  {journal} {CERN-ISR-PO-72-31}\ } (\bibinfo {year} {1972})}\BibitemShut {NoStop}%
\bibitem [{\citenamefont {Van~der Meer}(1995)}]{van1995stochastic}%
  \BibitemOpen
  \bibfield  {author} {\bibinfo {author} {\bibfnamefont {S.}~\bibnamefont {Van~der Meer}},\ }\bibfield  {title} {\bibinfo {title} {Stochastic damping of betatron oscillations},\ }\href@noop {} {\bibfield  {journal} {\bibinfo  {journal} {The Development of Colliders}\ ,\ \bibinfo {pages} {261}} (\bibinfo {year} {1995})}\BibitemShut {NoStop}%
\bibitem [{\citenamefont {Lebedev}(2024)}]{Lebedev_2024}%
  \BibitemOpen
  \bibfield  {author} {\bibinfo {author} {\bibfnamefont {V.}~\bibnamefont {Lebedev}},\ }\bibfield  {title} {\bibinfo {title} {High energy cooling},\ }\href {https://doi.org/10.1088/1748-0221/19/06/P06015} {\bibfield  {journal} {\bibinfo  {journal} {Journal of Instrumentation}\ }\textbf {\bibinfo {volume} {19}}\bibinfo  {number} { (06)},\ \bibinfo {pages} {P06015}}\BibitemShut {NoStop}%
\bibitem [{\citenamefont {Nagaitsev}\ \emph {et~al.}(2006)\citenamefont {Nagaitsev}, \citenamefont {Broemmelsiek}, \citenamefont {Burov}, \citenamefont {Carlson}, \citenamefont {Gattuso}, \citenamefont {Hu}, \citenamefont {Kroc}, \citenamefont {Prost}, \citenamefont {Pruss}, \citenamefont {Sutherland}, \citenamefont {Schmidt}, \citenamefont {Shemyakin}, \citenamefont {Tupikov}, \citenamefont {Warner}, \citenamefont {Kazakevich},\ and\ \citenamefont {Seletskiy}}]{ECool_PhysRevLett.96.044801}%
  \BibitemOpen
\bibfield  {number} {  }\bibfield  {author} {\bibinfo {author} {\bibfnamefont {S.}~\bibnamefont {Nagaitsev}}, \bibinfo {author} {\bibfnamefont {D.}~\bibnamefont {Broemmelsiek}}, \bibinfo {author} {\bibfnamefont {A.}~\bibnamefont {Burov}}, \bibinfo {author} {\bibfnamefont {K.}~\bibnamefont {Carlson}}, \bibinfo {author} {\bibfnamefont {C.}~\bibnamefont {Gattuso}}, \bibinfo {author} {\bibfnamefont {M.}~\bibnamefont {Hu}}, \bibinfo {author} {\bibfnamefont {T.}~\bibnamefont {Kroc}}, \bibinfo {author} {\bibfnamefont {L.}~\bibnamefont {Prost}}, \bibinfo {author} {\bibfnamefont {S.}~\bibnamefont {Pruss}}, \bibinfo {author} {\bibfnamefont {M.}~\bibnamefont {Sutherland}}, \bibinfo {author} {\bibfnamefont {C.~W.}\ \bibnamefont {Schmidt}}, \bibinfo {author} {\bibfnamefont {A.}~\bibnamefont {Shemyakin}}, \bibinfo {author} {\bibfnamefont {V.}~\bibnamefont {Tupikov}}, \bibinfo {author} {\bibfnamefont {A.}~\bibnamefont {Warner}}, \bibinfo {author} {\bibfnamefont {G.}~\bibnamefont {Kazakevich}},\ and\ \bibinfo {author}
  {\bibfnamefont {S.}~\bibnamefont {Seletskiy}},\ }\bibfield  {title} {\bibinfo {title} {Experimental demonstration of relativistic electron cooling},\ }\href {https://doi.org/10.1103/PhysRevLett.96.044801} {\bibfield  {journal} {\bibinfo  {journal} {Phys. Rev. Lett.}\ }\textbf {\bibinfo {volume} {96}},\ \bibinfo {pages} {044801} (\bibinfo {year} {2006})}\BibitemShut {NoStop}%
\bibitem [{\citenamefont {Lebedev}\ \emph {et~al.}(2021)\citenamefont {Lebedev}, \citenamefont {Jarvis}, \citenamefont {Piekarz}, \citenamefont {Romanov}, \citenamefont {Ruan},\ and\ \citenamefont {Andorf}}]{Lebedev_2021}%
  \BibitemOpen
  \bibfield  {author} {\bibinfo {author} {\bibfnamefont {V.}~\bibnamefont {Lebedev}}, \bibinfo {author} {\bibfnamefont {J.}~\bibnamefont {Jarvis}}, \bibinfo {author} {\bibfnamefont {H.}~\bibnamefont {Piekarz}}, \bibinfo {author} {\bibfnamefont {A.}~\bibnamefont {Romanov}}, \bibinfo {author} {\bibfnamefont {J.}~\bibnamefont {Ruan}},\ and\ \bibinfo {author} {\bibfnamefont {M.}~\bibnamefont {Andorf}},\ }\bibfield  {title} {\bibinfo {title} {The design of optical stochastic cooling for iota},\ }\href {https://doi.org/10.1088/1748-0221/16/05/T05002} {\bibfield  {journal} {\bibinfo  {journal} {Journal of Instrumentation}\ }\textbf {\bibinfo {volume} {16}}\bibinfo  {number} { (05)},\ \bibinfo {pages} {T05002}}\BibitemShut {NoStop}%
\bibitem [{\citenamefont {Zolotorev}\ and\ \citenamefont {Zholents}(1994)}]{OSCZolotarev}%
  \BibitemOpen
\bibfield  {number} {  }\bibfield  {author} {\bibinfo {author} {\bibfnamefont {M.~S.}\ \bibnamefont {Zolotorev}}\ and\ \bibinfo {author} {\bibfnamefont {A.~A.}\ \bibnamefont {Zholents}},\ }\bibfield  {title} {\bibinfo {title} {Transit-time method of optical stochastic cooling},\ }\href {https://doi.org/10.1103/PhysRevE.50.3087} {\bibfield  {journal} {\bibinfo  {journal} {Phys. Rev. E}\ }\textbf {\bibinfo {volume} {50}},\ \bibinfo {pages} {3087} (\bibinfo {year} {1994})}\BibitemShut {NoStop}%
\bibitem [{\citenamefont {Derbenev}(1992)}]{Derbenev_CeC_1992}%
  \BibitemOpen
  \bibfield  {author} {\bibinfo {author} {\bibfnamefont {Y.~S.}\ \bibnamefont {Derbenev}},\ }\bibfield  {title} {\bibinfo {title} {On possibilities of fast cooling of heavy particle beams},\ }\href {https://doi.org/10.1063/1.42152} {\bibfield  {journal} {\bibinfo  {journal} {AIP Conference Proceedings}\ }\textbf {\bibinfo {volume} {253}},\ \bibinfo {pages} {103} (\bibinfo {year} {1992})}\BibitemShut {NoStop}%
\bibitem [{\citenamefont {Litvinenko}\ and\ \citenamefont {Derbenev}(2009)}]{LitvinenkoCEC}%
  \BibitemOpen
  \bibfield  {author} {\bibinfo {author} {\bibfnamefont {V.~N.}\ \bibnamefont {Litvinenko}}\ and\ \bibinfo {author} {\bibfnamefont {Y.~S.}\ \bibnamefont {Derbenev}},\ }\bibfield  {title} {\bibinfo {title} {Coherent electron cooling},\ }\href {https://doi.org/10.1103/PhysRevLett.102.114801} {\bibfield  {journal} {\bibinfo  {journal} {Phys. Rev. Lett.}\ }\textbf {\bibinfo {volume} {102}},\ \bibinfo {pages} {114801} (\bibinfo {year} {2009})}\BibitemShut {NoStop}%
\bibitem [{\citenamefont {Ratner}(2013)}]{CECwithMicrobunchingAsAmplifier}%
  \BibitemOpen
  \bibfield  {author} {\bibinfo {author} {\bibfnamefont {D.}~\bibnamefont {Ratner}},\ }\bibfield  {title} {\bibinfo {title} {Microbunched electron cooling for high-energy hadron beams},\ }\href {https://doi.org/10.1103/PhysRevLett.111.084802} {\bibfield  {journal} {\bibinfo  {journal} {Phys. Rev. Lett.}\ }\textbf {\bibinfo {volume} {111}},\ \bibinfo {pages} {084802} (\bibinfo {year} {2013})}\BibitemShut {NoStop}%
\bibitem [{\citenamefont {Nagaitsev}\ \emph {et~al.}(2021)\citenamefont {Nagaitsev}, \citenamefont {Bergan}, \citenamefont {Lebedev}, \citenamefont {Stupakov},\ and\ \citenamefont {Wang}}]{nagaitsev:ipac2021-wepab273}%
  \BibitemOpen
  \bibfield  {author} {\bibinfo {author} {\bibfnamefont {S.}~\bibnamefont {Nagaitsev}}, \bibinfo {author} {\bibfnamefont {W.}~\bibnamefont {Bergan}}, \bibinfo {author} {\bibfnamefont {V.}~\bibnamefont {Lebedev}}, \bibinfo {author} {\bibfnamefont {G.}~\bibnamefont {Stupakov}},\ and\ \bibinfo {author} {\bibfnamefont {E.}~\bibnamefont {Wang}},\ }\bibfield  {title} {\bibinfo {title} {Cooling and diffusion rates in coherent electron cooling concepts},\ }in\ \href {https://doi.org/10.18429/JACoW-IPAC2021-WEPAB273} {\emph {\bibinfo {booktitle} {Proc. IPAC'21}}},\ \bibinfo {series and number} {\bibinfo {series} {International Particle Accelerator Conference}\ No.~\bibinfo {number} {12}}\ (\bibinfo  {publisher} {JACoW Publishing, Geneva, Switzerland},\ \bibinfo {year} {2021})\ pp.\ \bibinfo {pages} {3281--3284}\BibitemShut {NoStop}%
\bibitem [{\citenamefont {Willeke}\ \emph {et~al.}(2021)\citenamefont {Willeke} \emph {et~al.}}]{eicTechReport}%
  \BibitemOpen
  \bibfield  {author} {\bibinfo {author} {\bibfnamefont {F.}~\bibnamefont {Willeke}} \emph {et~al.},\ }\href {https://doi.org/10.2172/1765663} {\emph {\bibinfo {title} {Electron Ion Collider Conceptual Design Report}}}\ (\bibinfo {year} {2021})\BibitemShut {NoStop}%
\bibitem [{\citenamefont {Stupakov}(2018)}]{stupakov2018coolingShotNoiseEffect}%
  \BibitemOpen
  \bibfield  {author} {\bibinfo {author} {\bibfnamefont {G.}~\bibnamefont {Stupakov}},\ }\bibfield  {title} {\bibinfo {title} {Cooling rate for microbunched electron cooling without amplification},\ }\href {https://doi.org/10.1103/PhysRevAccelBeams.21.114402} {\bibfield  {journal} {\bibinfo  {journal} {Phys. Rev. Accel. Beams}\ }\textbf {\bibinfo {volume} {21}},\ \bibinfo {pages} {114402} (\bibinfo {year} {2018})}\BibitemShut {NoStop}%
\bibitem [{\citenamefont {Stupakov}(2010)}]{stupakov2010seededFELnoise}%
  \BibitemOpen
  \bibfield  {author} {\bibinfo {author} {\bibfnamefont {G.}~\bibnamefont {Stupakov}},\ }\href@noop {} {\emph {\bibinfo {title} {Noise amplification in hghg seeding}}},\ \bibinfo {type} {Tech. Rep.}\ (\bibinfo  {institution} {SLAC National Accelerator Lab., Menlo Park, CA (United States)},\ \bibinfo {year} {2010})\BibitemShut {NoStop}%
\bibitem [{\citenamefont {Litvinenko}(2009)}]{LitvinenkoFEL09}%
  \BibitemOpen
  \bibfield  {author} {\bibinfo {author} {\bibfnamefont {V.}~\bibnamefont {Litvinenko}},\ }\bibfield  {title} {\bibinfo {title} {Suppression of shot noise and spontaneous radiation in electron beams},\ }\href@noop {} {\bibfield  {journal} {\bibinfo  {journal} {Proceedings of FEL2009, Liverpool, UK}\ } (\bibinfo {year} {2009})}\BibitemShut {NoStop}%
\bibitem [{\citenamefont {Gover}\ \emph {et~al.}(2012)\citenamefont {Gover}, \citenamefont {Nause}, \citenamefont {Dyunin},\ and\ \citenamefont {Fedurin}}]{gover2012noiseLevelDecreaseInDriftSection}%
  \BibitemOpen
  \bibfield  {author} {\bibinfo {author} {\bibfnamefont {A.}~\bibnamefont {Gover}}, \bibinfo {author} {\bibfnamefont {A.}~\bibnamefont {Nause}}, \bibinfo {author} {\bibfnamefont {E.}~\bibnamefont {Dyunin}},\ and\ \bibinfo {author} {\bibfnamefont {M.}~\bibnamefont {Fedurin}},\ }\bibfield  {title} {\bibinfo {title} {Beating the shot-noise limit},\ }\href {https://doi.org/https://doi.org/10.1038/nphys2443} {\bibfield  {journal} {\bibinfo  {journal} {Nature Physics}\ }\textbf {\bibinfo {volume} {8}},\ \bibinfo {pages} {877} (\bibinfo {year} {2012})}\BibitemShut {NoStop}%
\bibitem [{\citenamefont {Huang}\ and\ \citenamefont {Stupakov}(2018)}]{HUANG2018182}%
  \BibitemOpen
  \bibfield  {author} {\bibinfo {author} {\bibfnamefont {Z.}~\bibnamefont {Huang}}\ and\ \bibinfo {author} {\bibfnamefont {G.}~\bibnamefont {Stupakov}},\ }\bibfield  {title} {\bibinfo {title} {Control and application of beam microbunching in high brightness linac-driven free electron lasers},\ }\href {https://doi.org/https://doi.org/10.1016/j.nima.2018.02.030} {\bibfield  {journal} {\bibinfo  {journal} {Nuclear Instruments and Methods in Physics Research Section A: Accelerators, Spectrometers, Detectors and Associated Equipment}\ }\textbf {\bibinfo {volume} {907}},\ \bibinfo {pages} {182} (\bibinfo {year} {2018})}\BibitemShut {NoStop}%
\bibitem [{\citenamefont {Loos}\ \emph {et~al.}(2008)\citenamefont {Loos}, \citenamefont {Akre}, \citenamefont {Brachmann}, \citenamefont {Decker}, \citenamefont {Ding}, \citenamefont {Dowell}, \citenamefont {Emma}, \citenamefont {Frisch}, \citenamefont {Gilevich}, \citenamefont {Hays}, \citenamefont {Hering}, \citenamefont {Huang}, \citenamefont {Iverson}, \citenamefont {Limborg}, \citenamefont {Miahnahri}, \citenamefont {Molloy}, \citenamefont {Nuhn}, \citenamefont {Turner}, \citenamefont {Welch},\ and\ \citenamefont {Menlo}}]{LoosLCLSCOTR}%
  \BibitemOpen
  \bibfield  {author} {\bibinfo {author} {\bibfnamefont {H.}~\bibnamefont {Loos}}, \bibinfo {author} {\bibfnamefont {R.}~\bibnamefont {Akre}}, \bibinfo {author} {\bibfnamefont {A.}~\bibnamefont {Brachmann}}, \bibinfo {author} {\bibfnamefont {F.-J.}\ \bibnamefont {Decker}}, \bibinfo {author} {\bibfnamefont {Y.}~\bibnamefont {Ding}}, \bibinfo {author} {\bibfnamefont {D.}~\bibnamefont {Dowell}}, \bibinfo {author} {\bibfnamefont {P.}~\bibnamefont {Emma}}, \bibinfo {author} {\bibfnamefont {J.}~\bibnamefont {Frisch}}, \bibinfo {author} {\bibfnamefont {S.}~\bibnamefont {Gilevich}}, \bibinfo {author} {\bibfnamefont {G.}~\bibnamefont {Hays}}, \bibinfo {author} {\bibfnamefont {P.}~\bibnamefont {Hering}}, \bibinfo {author} {\bibfnamefont {Z.}~\bibnamefont {Huang}}, \bibinfo {author} {\bibfnamefont {R.}~\bibnamefont {Iverson}}, \bibinfo {author} {\bibfnamefont {C.}~\bibnamefont {Limborg}}, \bibinfo {author} {\bibfnamefont {A.}~\bibnamefont {Miahnahri}}, \bibinfo {author} {\bibfnamefont {S.}~\bibnamefont {Molloy}},
  \bibinfo {author} {\bibfnamefont {H.-D.}\ \bibnamefont {Nuhn}}, \bibinfo {author} {\bibfnamefont {J.}~\bibnamefont {Turner}}, \bibinfo {author} {\bibfnamefont {J.}~\bibnamefont {Welch}},\ and\ \bibinfo {author} {\bibfnamefont {S.}~\bibnamefont {Menlo}},\ }\bibfield  {title} {\bibinfo {title} {Observation of coherent optical transition radiation in the lcls linac},\ }\href@noop {} {\bibfield  {journal} {\bibinfo  {journal} {30th International Free Electron Laser Conference, FEL 2008}\ } (\bibinfo {year} {2008})}\BibitemShut {NoStop}%
\bibitem [{\citenamefont {Nagaitsev}\ \emph {et~al.}(2022)\citenamefont {Nagaitsev} \emph {et~al.}}]{nagaitsev:napac2022-mopa34}%
  \BibitemOpen
  \bibfield  {author} {\bibinfo {author} {\bibfnamefont {S.}~\bibnamefont {Nagaitsev}} \emph {et~al.},\ }\bibfield  {title} {\bibinfo {title} {{Noise in Intense Electron Bunches}},\ }in\ \href {https://doi.org/10.18429/JACoW-NAPAC2022-MOPA34} {\emph {\bibinfo {booktitle} {Proc. 5th Int. Particle Accel. Conf. (NAPAC'22)}}},\ \bibinfo {series and number} {\bibinfo {series} {International Particle Accelerator Conference}\ No.~\bibinfo {number} {5}}\ (\bibinfo  {publisher} {JACoW Publishing, Geneva, Switzerland},\ \bibinfo {year} {2022})\ pp.\ \bibinfo {pages} {128--131}\BibitemShut {NoStop}%
\bibitem [{\citenamefont {Wiedemann}(2015)}]{wiedemann}%
  \BibitemOpen
  \bibfield  {author} {\bibinfo {author} {\bibfnamefont {H.}~\bibnamefont {Wiedemann}},\ }\href {https://doi.org/10.1007/978-3-319-18317-6} {\emph {\bibinfo {title} {Particle Accelerator Physics}}},\ \bibinfo {edition} {4th}\ ed.\ (\bibinfo  {publisher} {Springer Cham},\ \bibinfo {year} {2015})\BibitemShut {NoStop}%
\bibitem [{\citenamefont {Landau}\ and\ \citenamefont {Lifshitz}(1984)}]{LANDAU1984394}%
  \BibitemOpen
  \bibfield  {author} {\bibinfo {author} {\bibfnamefont {L.}~\bibnamefont {Landau}}\ and\ \bibinfo {author} {\bibfnamefont {E.}~\bibnamefont {Lifshitz}},\ }\bibfield  {title} {\bibinfo {title} {Chapter {XIV} - {T}he {P}assage of {F}ast {P}articles {T}hrough {M}atter},\ }in\ \href {https://doi.org/https://doi.org/10.1016/B978-0-08-030275-1.50020-5} {\emph {\bibinfo {booktitle} {Electrodynamics of Continuous Media (Second Edition)}}},\ \bibinfo {series} {Course of Theoretical Physics}, Vol.~\bibinfo {volume} {8}\ (\bibinfo  {publisher} {Pergamon},\ \bibinfo {address} {Amsterdam},\ \bibinfo {year} {1984})\ \bibinfo {edition} {second edition}\ ed.,\ pp.\ \bibinfo {pages} {394--412}\BibitemShut {NoStop}%
\bibitem [{\citenamefont {Fano}(1947)}]{PhysRev.72.26}%
  \BibitemOpen
  \bibfield  {author} {\bibinfo {author} {\bibfnamefont {U.}~\bibnamefont {Fano}},\ }\bibfield  {title} {\bibinfo {title} {Ionization yield of radiations. ii. the fluctuations of the number of ions},\ }\href {https://doi.org/10.1103/PhysRev.72.26} {\bibfield  {journal} {\bibinfo  {journal} {Phys. Rev.}\ }\textbf {\bibinfo {volume} {72}},\ \bibinfo {pages} {26} (\bibinfo {year} {1947})}\BibitemShut {NoStop}%
\bibitem [{\citenamefont {Beenakker}\ and\ \citenamefont {Schönenberger}(2003)}]{Beenakker_quantumShotNoise}%
  \BibitemOpen
  \bibfield  {author} {\bibinfo {author} {\bibfnamefont {C.}~\bibnamefont {Beenakker}}\ and\ \bibinfo {author} {\bibfnamefont {C.}~\bibnamefont {Schönenberger}},\ }\bibfield  {title} {\bibinfo {title} {Quantum shot noise},\ }\href {https://doi.org/10.1063/1.1583532} {\bibfield  {journal} {\bibinfo  {journal} {Physics Today}\ }\textbf {\bibinfo {volume} {56}},\ \bibinfo {pages} {37} (\bibinfo {year} {2003})}\BibitemShut {NoStop}%
\bibitem [{\citenamefont {Orlandi}(2006)}]{orlandi2006transvSizeEffects}%
  \BibitemOpen
  \bibfield  {author} {\bibinfo {author} {\bibfnamefont {G.~L.}\ \bibnamefont {Orlandi}},\ }\bibfield  {title} {\bibinfo {title} {Spatial coherence in the transition radiation spectrum},\ }\href {https://doi.org/https://doi.org/10.1016/j.optcom.2006.06.056} {\bibfield  {journal} {\bibinfo  {journal} {Optics Communications}\ }\textbf {\bibinfo {volume} {267}},\ \bibinfo {pages} {322} (\bibinfo {year} {2006})}\BibitemShut {NoStop}%
\bibitem [{\citenamefont {Chiadroni}\ \emph {et~al.}(2012)\citenamefont {Chiadroni}, \citenamefont {Castellano}, \citenamefont {Cianchi}, \citenamefont {Honkavaara},\ and\ \citenamefont {Kube}}]{chiadroni2012TransvSizeEffects}%
  \BibitemOpen
  \bibfield  {author} {\bibinfo {author} {\bibfnamefont {E.}~\bibnamefont {Chiadroni}}, \bibinfo {author} {\bibfnamefont {M.}~\bibnamefont {Castellano}}, \bibinfo {author} {\bibfnamefont {A.}~\bibnamefont {Cianchi}}, \bibinfo {author} {\bibfnamefont {K.}~\bibnamefont {Honkavaara}},\ and\ \bibinfo {author} {\bibfnamefont {G.}~\bibnamefont {Kube}},\ }\bibfield  {title} {\bibinfo {title} {Effects of transverse electron beam size on transition radiation angular distribution},\ }\href {https://doi.org/https://doi.org/10.1016/j.nima.2012.01.011} {\bibfield  {journal} {\bibinfo  {journal} {Nuclear Instruments and Methods in Physics Research Section A: Accelerators, Spectrometers, Detectors and Associated Equipment}\ }\textbf {\bibinfo {volume} {673}},\ \bibinfo {pages} {56} (\bibinfo {year} {2012})}\BibitemShut {NoStop}%
\bibitem [{\citenamefont {Lumpkin}\ \emph {et~al.}(2009)\citenamefont {Lumpkin}, \citenamefont {Dejus},\ and\ \citenamefont {Sereno}}]{lumpkinCompressionInFEL}%
  \BibitemOpen
  \bibfield  {author} {\bibinfo {author} {\bibfnamefont {A.~H.}\ \bibnamefont {Lumpkin}}, \bibinfo {author} {\bibfnamefont {R.~J.}\ \bibnamefont {Dejus}},\ and\ \bibinfo {author} {\bibfnamefont {N.~S.}\ \bibnamefont {Sereno}},\ }\bibfield  {title} {\bibinfo {title} {Coherent optical transition radiation and self-amplified spontaneous emission generated by chicane-compressed electron beams},\ }\href {https://doi.org/10.1103/PhysRevSTAB.12.040704} {\bibfield  {journal} {\bibinfo  {journal} {Phys. Rev. ST Accel. Beams}\ }\textbf {\bibinfo {volume} {12}},\ \bibinfo {pages} {040704} (\bibinfo {year} {2009})}\BibitemShut {NoStop}%
\bibitem [{\citenamefont {Tremaine}\ \emph {et~al.}(1998)\citenamefont {Tremaine}, \citenamefont {Rosenzweig}, \citenamefont {Anderson}, \citenamefont {Frigola}, \citenamefont {Hogan}, \citenamefont {Murokh}, \citenamefont {Pellegrini}, \citenamefont {Nguyen},\ and\ \citenamefont {Sheffield}}]{SASEfelMicrobunching}%
  \BibitemOpen
  \bibfield  {author} {\bibinfo {author} {\bibfnamefont {A.}~\bibnamefont {Tremaine}}, \bibinfo {author} {\bibfnamefont {J.~B.}\ \bibnamefont {Rosenzweig}}, \bibinfo {author} {\bibfnamefont {S.}~\bibnamefont {Anderson}}, \bibinfo {author} {\bibfnamefont {P.}~\bibnamefont {Frigola}}, \bibinfo {author} {\bibfnamefont {M.}~\bibnamefont {Hogan}}, \bibinfo {author} {\bibfnamefont {A.}~\bibnamefont {Murokh}}, \bibinfo {author} {\bibfnamefont {C.}~\bibnamefont {Pellegrini}}, \bibinfo {author} {\bibfnamefont {D.~C.}\ \bibnamefont {Nguyen}},\ and\ \bibinfo {author} {\bibfnamefont {R.~L.}\ \bibnamefont {Sheffield}},\ }\bibfield  {title} {\bibinfo {title} {Observation of self-amplified spontaneous-emission-induced electron-beam microbunching using coherent transition radiation},\ }\href {https://doi.org/10.1103/PhysRevLett.81.5816} {\bibfield  {journal} {\bibinfo  {journal} {Phys. Rev. Lett.}\ }\textbf {\bibinfo {volume} {81}},\ \bibinfo {pages} {5816} (\bibinfo {year} {1998})}\BibitemShut {NoStop}%
\bibitem [{\citenamefont {Wesch}\ \emph {et~al.}(2009)\citenamefont {Wesch}, \citenamefont {Behrens}, \citenamefont {Schmidt},\ and\ \citenamefont {Schm{\"u}ser}}]{weschFLASHOTR}%
  \BibitemOpen
  \bibfield  {author} {\bibinfo {author} {\bibfnamefont {S.}~\bibnamefont {Wesch}}, \bibinfo {author} {\bibfnamefont {C.}~\bibnamefont {Behrens}}, \bibinfo {author} {\bibfnamefont {B.}~\bibnamefont {Schmidt}},\ and\ \bibinfo {author} {\bibfnamefont {P.}~\bibnamefont {Schm{\"u}ser}},\ }\bibfield  {title} {\bibinfo {title} {Observation of coherent optical transition radiation and evidence for microbunching in magnetic chicanes},\ }in\ \href@noop {} {\emph {\bibinfo {booktitle} {Proc. 31st Int. Free Electron Laser Conf.(FEL’09)}}}\ (\bibinfo {year} {2009})\ pp.\ \bibinfo {pages} {619--622}\BibitemShut {NoStop}%
\bibitem [{\citenamefont {Floettmann}(2000)}]{ASTRA}%
  \BibitemOpen
  \bibfield  {author} {\bibinfo {author} {\bibfnamefont {K.}~\bibnamefont {Floettmann}},\ }\href@noop {} {\bibinfo {title} {Astra}},\ \bibinfo {howpublished} {\url{https://github.com/NIUaard/FAST/tree/master}} (\bibinfo {year} {2000})\BibitemShut {NoStop}%
\bibitem [{\citenamefont {Huebl}\ \emph {et~al.}(2022)\citenamefont {Huebl} \emph {et~al.}}]{impactX}%
  \BibitemOpen
  \bibfield  {author} {\bibinfo {author} {\bibfnamefont {A.}~\bibnamefont {Huebl}} \emph {et~al.},\ }\bibfield  {title} {\bibinfo {title} {Next generation computational tools for the modeling and design of particle accelerators at exascale},\ }\href@noop {} {\bibfield  {journal} {\bibinfo  {journal} {arXiv preprint arXiv:2208.02382}\ } (\bibinfo {year} {2022})}\BibitemShut {NoStop}%
\bibitem [{FAS()}]{FASTlatticeGitHub}%
  \BibitemOpen
  \href@noop {} {\bibinfo {title} {Input decks for elegant, astra to simulate the fast electron-linac beamline}},\ \bibinfo {howpublished} {\url{https://github.com/NIUaard/FAST/tree/master}}\BibitemShut {NoStop}%
\end{thebibliography}
\end{document}